\documentclass[
    aps, prb,
    superscriptaddress, 
    showkeywords,
    reprint,
    longbibliography,
    amsmath,amssymb,floatfix.xcolor=table
    ]{revtex4-2}

\usepackage{url}

\usepackage[utf8]{inputenc}
\usepackage{ulem}
\usepackage{amsmath}
\usepackage{amssymb}

\usepackage[T1]{fontenc} 
\usepackage[LGR,T1]{fontenc}

\usepackage{textgreek}
\DeclareUnicodeCharacter{03B4}{\textdelta}
\usepackage{graphicx}
\usepackage[separate-uncertainty=true]{siunitx}
\usepackage{newunicodechar}
\newunicodechar{̈}{\"{}}
\usepackage{textcomp}

\usepackage{xcolor}

\usepackage{hyperref}

\DeclareUnicodeCharacter{2215}{\\} 

\makeatletter
\def\maketitle{
	\@author@finish
	\title@column\titleblock@produce
	\suppressfloats[t]}
\makeatother

\newcommand{\beginsupplement}{
		\setcounter{page}{1}
		\setcounter{section}{0}
		\renewcommand{\thesection}{S\arabic{section}}%
		\setcounter{table}{0}
		\renewcommand{\thetable}{S\arabic{table}}%
		\setcounter{figure}{0}
		\renewcommand{\thefigure}{S\arabic{figure}}%
		\setcounter{equation}{0}
		\renewcommand{\theequation}{S\arabic{equation}}%
		}

\newcommand{\iitm}{Department of Physics, Indian Institute of Technology Madras, Chennai 600036, India}
\newcommand{\forg}{Functional Oxides Research Group, Department of Physics, Indian Institute of Technology Madras, Chennai 600036, India}
\newcommand{\iitd}{Department of Physics, Indian Institute of Technology Delhi, Hauz Khas, New Delhi 110016, India}

\newcommand{\papertitle}{Controlling Spin-Mixing Conductance in KTaO$_{3}$ 2DEGs by Varying Argon-Ion Irradiation Time}

\begin{document}
	
	\title{\papertitle}
    \author{Yasar K. Arafath}
		\affiliation{\iitm}

        \author{Vaishali Yadav}
		\affiliation{\iitd}

        \author{Nidhi Kandwal}
		\affiliation{\iitd}        

        \author{P.N. Santhosh}
		\affiliation{\iitm}
        \affiliation{\forg}

         \author{Pranaba Kishore Muduli}
		\affiliation{\iitd}
	
	\author{Prasanta Kumar Muduli}
	       \email{muduli@iitm.ac.in}
		    \affiliation{\iitm}
	
	\begin{abstract}
		
		The Rashba-split two-dimensional electron gas (2DEG) at the surface and interface of insulating oxides like KTaO$_{3}$ (KTO) shows great promise for all-oxide spintronics. However, efficient spin current injection into the adjacent 2DEG remains a key challenge.  In this study, we report the spin-pumping experiments on a 2DEG formed on the (001)KTO surface via Ar$^+$ irradiation. We observed a significant increase in magnetic damping in the Ar$^+$-KTO/Py bilayer compared to a non-irradiated KTO/Py control sample, confirming spin pumping into the 2DEG. We demonstrate that the spin-mixing conductance ($g_{\uparrow\downarrow}^r$) can be substantially enhanced by controlling the Ar$^+$ irradiation time. The enhancement is attributed to increased 2DEG conductance, which results from a higher concentration of oxygen vacancies with longer irradiation times. This work provides crucial guidance for optimizing spin-to-charge conversion in KTO-based systems, highlighting the potential of Ar$^+$-irradiated KTO 2DEGs for future oxide spintronics.
		
	\end{abstract}
	
	
\maketitle
	
\section{Introduction}
    
The study of two-dimensional electron gases (2DEGs) at complex oxide surfaces and interfaces has received considerable attention since their 2004 discovery at the interface between two well-known perovskite insulators, LaAlO$_3$ (LAO) and SrTiO$_3$ (STO) \cite{Ohtomo2004,Hwang2012,Chen2023,Varignon2018}. The strong electronic correlations in oxide 2DEGs give rise to a remarkable range of emergent properties, including interfacial magnetism \cite{Dikin2011, Li2011}, two dimensional superconductivity\cite{Reyren2007}, multiferroicity \cite{Brhin2023}, Shubnikov-de Haas oscillations\cite{Rubi2020}, quantum Hall effect \cite{Matsubara2016}, persistent photoconductivity \cite{Tebano2012}, and gate-tunable Rashba effect \cite{Caviglia2008, Lesne2016}. Besides the metallic behavior, the oxide 2DEG has very high mobilities $\sim 10^{4} \mathrm{~cm}^{2} / V s$ with low sheet carrier densities $\sim 10^{14} \mathrm{~cm}^{-2}$, which makes them ideal candidates for developing next-generation all-oxide devices \cite{Herranz2007}. Over the past years, a variety of techniques have been developed to create oxide 2DEGs, such as electrostatic gating \cite{Gallagher2015,Zhang2017}, metallic capping\cite{Rdel2016, Vicente-Arche2021}, ultraviolet light exposure~\cite{Meevasana2011}, and $\mathrm{Ar}^{+}$-irradiation \cite{Reagor2005}. Among these $\mathrm{Ar}^{+}$-irradiation stands out as a simple and fast
method for creating a 2DEG on the oxide surface. This technique is particularly valuable because the 2DEG's key properties-thickness, resistivity, carrier density, and mobility-can be precisely adjusted by controlling the $\mathrm{Ar}^{+}$ion's energy and exposure time \cite{Chang2015}. $\mathrm{Ar}^{+}$irradiation creates a metallic 2DEG by inducing oxygen vacancies on the oxide surface \cite{QiuruWang2016,Santander-Syro2011-mw}. These vacancies act as electron donors, effectively doping the crystal surface with free electrons.

While most early research focused on STO-based systems, a new class of 2DEGs was discovered in 2012 based on the 5d transition metal oxide $\mathrm{KTaO}_{3}$ (KTO) \cite{Gupta2022}. Similar to STO, KTO is a wide-band-gap perovskite insulator, but its heavier tantalum (Ta) atom gives it a significantly stronger spin-orbit coupling, leading to more fascinating transport properties. High-mobility 2DEGs have been successfully created at interfaces between KTO and other oxides such as $\mathrm{LaAlO}_{3}$\cite{Zhang2017}, $\gamma-\mathrm{Al}_{2} \mathrm{O}_{3}$ \cite{Zhang2023}, LaVO$_3$ \cite{Wadehra2020}, CaZrO$_3$ \cite{Qi2024}, LaTiO$_3$\cite{Zou2015}, and EuO \cite{Hua2022,Liu2021}. Other fabrication methods for KTO-based 2DEGs include liquid and solid-state gating\cite{Nakamura2009,Ueno2011}, and $\mathrm{Ar}^{+}$irradiation \cite{Harashima2013}.

In oxide 2DEGs, broken inversion symmetry yields a significant Rashba-type spin-orbit coupling (SOC), whose momentum-dependent spin splitting enables efficient room temperature charge-to-spin interconversion through the Rashba-Edelstein effect (REE) and its inverse (IREE) \cite{Han2024}. The first demonstration of spin-to-charge conversion via the IREE in oxide 2DEGs was achieved at the LAO/STO interface in 2016 \cite{Lesne2016}. To date, spin-charge conversion in KTO-based 2DEGs has been reported in only four publications. The first report, in 2019 by Zhang et al. \cite{Zhang2019}, demonstrated spin-to-charge conversion via the IREE at the EuO/KTO interface, utilizing thermal spin current injection from EuO. Following this, Vicente-Arche et al. \cite{VicenteArche2021}, demonstrated high-efficiency spin-to-charge interconversion in (001)KTO/AlO$_x$ 2DEGs using a combination of spin-pumping ferromagnetic resonance (FMR) and unidirectional magnetoresistance. Further, Hui Zhang et al. \cite{Zhang2023} utilized spin-torque FMR to demonstrate charge-to-spin conversion in $\gamma-\text{Al}_2\text{O}_3/(\mathbf{001})\text{KTO}$, observing a spin-orbit torque efficiency an order of magnitude higher than that of heavy metals. Most recently, Athby H. Al-Tawhid et al. \cite{Al-Tawhid2025} studied spin-to-charge conversion in a $(\mathbf{111})\text{KTO}/\text{AlO}_x$ 2DEGs via spin pumping, revealing a non-trivial angle dependence. Although the efficiencies in KTO-based 2DEGs are an order of magnitude greater than those of transition metals like Pt, they still lag behind the values measured in STO-based 2DEGs.
This indicates that a substantial effort is required to realize KTO’s predicted theoretical advantage over STO. Crucially, no spin-to-charge interconversion experiments have used $\text{Ar}^+$ irradiation-prepared KTO 2DEGs, which is the focus of this paper

The spin-pumping method is a widely used technique to investigate spin-to-charge conversion in oxide 2DEGs~\cite{Zheng2024}. This technique uses a bilayer of a ferromagnetic metal (FM) and a non-magnetic metal (NM), in this case, the 2DEG. The spin pumping process involves the excitation of magnetization precession in the FM forced by a radio frequency (RF) microwave field. When the ferromagnetic resonance (FMR) condition for FM is fulfilled, a spin current is injected into the NM (here 2DEG) due to spin angular momentum conservation \cite{Saitoh2006,Mosendz2010}. A key signature of this spin injection is the increased Gilbert damping of the FM, which results from the loss of spin angular momentum to the neighboring NM (2DEG) layer \cite{Swindells2022}. This damping effect is more pronounced in NM layers with strong SOC, as it facilitates a more efficient transfer of spin. The most crucial parameters that quantify the efficiency of spin-to-charge current conversion in an FM/NM bilayer are: the spin-diffusion length ( $\lambda_{s}$ ), the spin Hall angle $\left(\theta_{S H}\right)$ of the NM and an interfacial parameter called the spin-mixing conductance $\left(g_{\uparrow \downarrow}\right)$. The real part of the spin-mixing conductance $g_{\uparrow \downarrow}^r$ is a critical parameter that quantifies how well spins can cross from the FM into the 2DEG\cite{Tserkovnyak2005}. For high conversion efficiency, a large $g_{\uparrow \downarrow}^r$ is crucial. In traditional epitaxially grown 2DEGs, such as LAO/STO, the spin current must pass through an insulating layer before reaching the 2DEG, which can limit this spin-mixing conductance \cite{Vicente-Arche2021}. In contrast, 2DEGs fabricated by $\mathrm{Ar}^{+}$irradiation offer a significant advantage: the FM layer can be deposited directly on the 2DEG surface. This direct contact maximizes the interfacial spin-mixing conductance, leading to more efficient spin-to-charge conversion and a stronger FMR signal.

In this paper, we present FMR-based spin-pumping experiments on $\mathrm{KTO}(001)$ surfaces where a 2DEG was created via $\mathrm{Ar}^{+}$irradiation. We chose the $\mathrm{KTO}(001)$ surface due to its higher spin-splitting compared to the (111) and (110) orientations. We observed a notable enhancement of magnetic damping in the $\mathrm{Ar}^{+}-\mathrm{KTO} / \mathrm{Py}$ bilayer relative to a control sample (KTO/Py without irradiation), which serves as definitive evidence of spin current injection into the 2DEG. Furthermore, we demonstrate that the efficiency of this spin transfer, quantified by the spin-mixing conductance $\left(g_{\uparrow \downarrow}^r\right)$, can be substantially enhanced by simply adjusting the $\mathrm{Ar}^{+}$irradiation time. We believe this enhancement is a result of increased 2DEG conductance with longer irradiation times. Given the limited research in this area, our findings offer key insights for optimizing spin-mixing conductance and spin-to-charge conversion in KTO 2DEGs, highlighting their potential for next-generation spin-orbitronic devices \cite{Vaz2018}.

\section{Experimental methods}
In our experiments, commercially available one-sided polished single-crystal (001)KTO substrates were used. The crystalline quality of the as-received substrates was verified using X-ray diffraction (XRD), as shown in the supplementary material (Section S1). Before irradiation, the substrates were cut into $3 \times 3 \mathrm{~mm}^{2}$ pieces. They were then cleaned by sonicating sequentially in acetone and isopropanol for one minute each, followed by drying with nitrogen gas.

The substrates were subsequently irradiated with a Kaufmann-type ion source at an acceleration voltage of 300 V and normal incidence. The base pressure of the vacuum chamber was maintained below $5 \times 10^{-6}$ Torr, with the Ar gas pressure during irradiation at $\approx 2$ mTorr. The irradiation time ( $t_{\mathrm{Ar}^{+}}$ ) ranged from 5 to 40 minutes on a water-cooled sample holder to prevent significant heating. For irradiation times exceeding 10 minutes, the process was conducted in 10 -minute intervals with 5 -minute pauses.

Two identical $3 \mathrm{~mm} \times 3 \mathrm{~mm}$ samples were irradiated at the same time. One sample was reserved for surface characterization, while the other was used for FMR measurements. The surface morphology of the irradiated substrates was analyzed using a Bruker atomic force microscope (AFM). X-ray photoelectron spectroscopy (XPS) measurements were performed at room temperature using a Thermoscientific system equipped with a $Al~  K_{\alpha}$ source. The XPS spectra were calibrated against the $\mathrm{C}~1 s$ signal from adsorbed carbon at a binding energy of 284.8 eV.

Immediately after $\mathrm{Ar}^{+}$ion irradiation, the KTO substrates were transferred to an ultra-high vacuum (UHV) sputter chamber where a 15 nm -thick $\mathrm{Ni}_{80} \mathrm{Fe}_{20}$ (Py) film was deposited {at rate $0.012$ nm/s} via DC sputtering. The sputtering was performed at room temperature under an Ar gas pressure of 2 mTorr and a power of 60 W. The base pressure of the chamber remained below $3 \times 10^{-7} \mathrm{mbar}$. Before deposition, the Py target was pre-sputtered for 2 minutes to ensure film purity. Following the Py deposition, a 2 nm-thick aluminum (Al) layer was deposited in-situ as a protective capping layer. This Al layer fully oxidizes upon exposure to air, forming an $\mathrm{AlO}_{\mathrm{x}}$ protective layer that prevents the Py layer from oxidizing. A series of bilayer samples, abbreviated as $\mathrm{Ar}^{+}-\mathrm{KTO} / \mathrm{Py}$, were prepared with varying $\mathrm{Ar}^{+}$ irradiation times. A reference KTO/Py sample was also prepared on a non-irradiated KTO substrate using the same deposition method.

FMR measurements were performed using a custom-made broadband spectrometer \cite{Bansal2018,Akash-prb-2018} with a coplanar waveguide (CPW) and a lock-in amplifier. The measurements were conducted by sweeping the DC magnetic field around the resonance field while keeping the microwave excitation frequency constant. The excitation frequency ranged from 4 to 10 GHz. The sample was placed with the film side down on the ground-signal-ground line of the CPW. All measurements were performed with an in-plane external
magnetic field. The electrical resistance measurements were carried out using a Quantum Design Dynacool PPMS system.

\section{Results and discussion}
\subsection{2DEG Fabrication and Surface Properties}

\begin{figure*}[htb]
    \centering
    \includegraphics[width=0.8\linewidth]{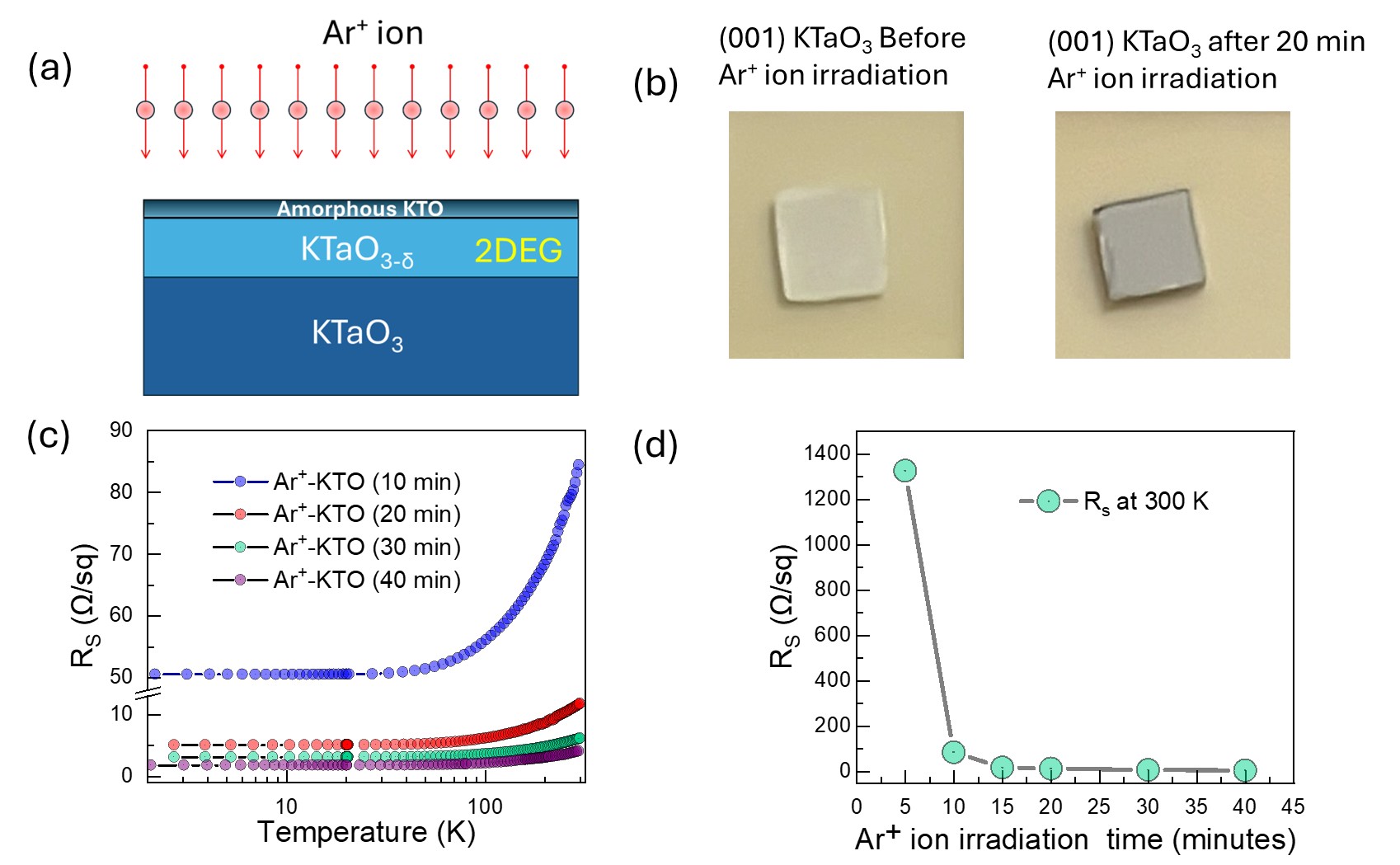}
    \caption{(a) Schematic of the creation of 2DEG on a single-crystal KTO surface via $\mathrm{Ar}^+$  ion irradiation. The $\mathrm{Ar}^+$  ion irradiation generates oxygen vacancies, forming a metallic 2DEG layer just beneath the resulting amorphous KTO top surface. (b) Photographs of the KTO substrate before and after 20 min of $\mathrm{Ar}^+$ ion irradiation, showing a visual change. (c) Sheet resistance ($R_S$) as a function of temperature for samples irradiated for different durations. The metallic behavior across all temperatures, without a low-temperature resistance upturn, confirms successful 2DEG formation. (d) Dependence of room-temperature sheet resistance ($R_S$) on $\mathrm{Ar}^+$ ion irradiation time, illustrating a rapid decrease in resistance as irradiation time increases.}
    \label{fig1}
\end{figure*}

Argon ion ( $\mathrm{Ar}^+$ ) irradiation is a straightforward and scalable method for creating a 2DEG on the surface of perovskite materials such as $\mathrm{KTO}$. This process generates a few-nanometer-thick conducting layer at the substrate surface as shown in Fig.~\ref{fig1}(a). Four samples were prepared with varying $\mathrm{Ar}^+$ irradiation durations, from 5 to 20 minutes, while other experimental parameters were kept constant. Following the irradiation, the KTO single crystals, which were originally white, underwent a color change to a grayish-black hue, as shown in Fig.~\ref{fig1}(b). Previous studies on STO substrates have shown that $\mathrm{Ar}^+$ irradiation damages the top few atomic layers, transforming them into an amorphous, insulating state ~\cite{Henrich1978, Wakabayashi2025,Gross2011}. The conducting 2DEG forms in the underlying crystalline layer, which becomes oxygen-deficient~\cite{QiuruWang2016, Psiuk2016}. The thickness of this amorphous layer remains largely constant, regardless of the irradiation time~\cite{Zhang2016,Wadehra2017}. We estimated the $\mathrm{Ar}^+$ irradiation penetration depth (proportional to the thickness of the 2DEG), $\mathrm{L}$, in angstroms using the following formula ~\cite{Reagor2005}:

\begin{equation}
L=1.1 \frac{E^{2 / 3} W_{K T O}}{\rho_{KTO}\left(Z_{A r}^{\frac{1}{4}}+Z_{K T O}^{\frac{1}{4}}\right)^{2}}  \label{eq1}
\end{equation}

Here, $E$ is the $\mathrm{Ar}^+$ ion energy in $\mathrm{eV}, W$ is the atomic weight of KTO (in atomic mass units), $\rho$ is the density of KTO, and $Z_{A r^{+}}$and $Z_{K T O}$ are the atomic numbers of the $\mathrm{Ar}^{+}$ion and KTO, respectively. Our calculations yielded a value of $\mathrm{L}=6.6\mathrm{~nm}$, suggesting a 2DEG thickness of approximately 6 nm. This is consistent with previous research on STO-based 2DEGs fabricated using similar $\mathrm{Ar}^{+}$ irradiation techniques, where$~\mathrm{Ar}^+$ ion irradiation resulted in a penetration depth of $\mathrm{L}=3.5$ to 12 nm \cite{DKumar2015,Schultz2007}.

\begin{figure*}[htb]
    \centering
    \includegraphics[width=0.8\linewidth]{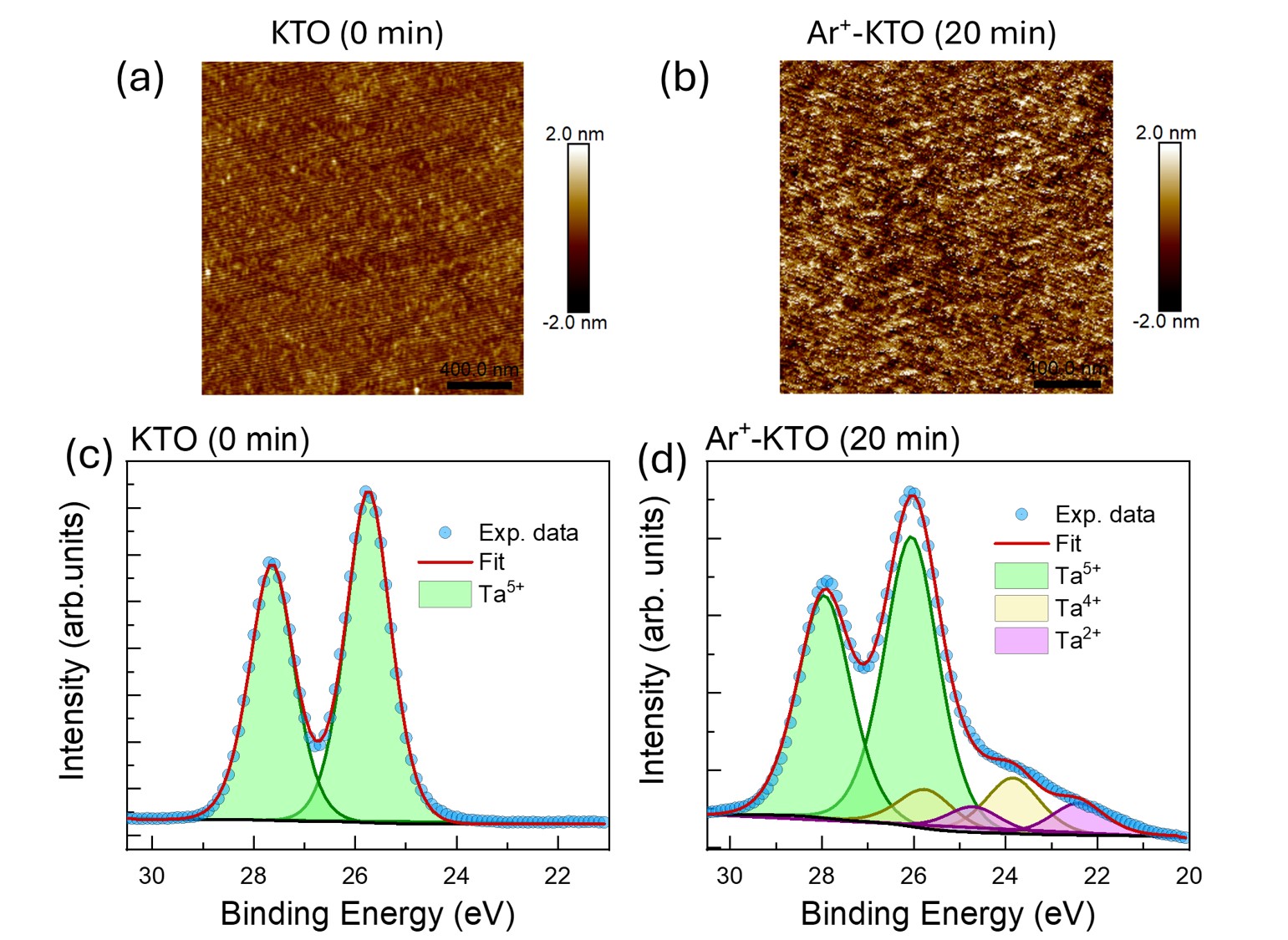}
    \caption{(a) Atomic Force Microscopy (AFM) image of the single-crystal KTO substrate before $\mathrm{Ar}^+$  ion irradiation, showing the initial surface morphology. (b) AFM image of the KTO substrate after 20 minutes of $\mathrm{Ar}^+$  ion irradiation, illustrating the increase in surface roughness. (c) High-resolution X-ray Photoelectron Spectroscopy (XPS) spectra of the Ta 4f core levels and associated peak fits for the KTO substrate before irradiation. (d) XPS spectra and fits of the Ta 4f core levels for the KTO substrate after 20 minutes of $\mathrm{Ar}^{+}$ion irradiation. The appearance of a new shoulder near 24 eV in the Ta 4 f spectrum confirms the presence of reduced tantalum species ( $\mathrm{Ta}^{4+}, \mathrm{Ta}^{2+}$ ), which is direct evidence for the formation of oxygen vacancies induced by the $\mathrm{Ar}^{+}$ irradiation.}
    \label{fig2}
\end{figure*}

To confirm the metallic nature of the 2DEG formed by $\mathrm{Ar}^+$ ion irradiation, we measured the temperature-dependent resistivity of all prepared samples. Using a conventional four-probe method in a Van der Pauw configuration, we measured the sheet resistance of the $\mathrm{Ar}^{+}$-irradiated KTO samples. As shown in Fig.~\ref {fig1}(c), all samples exhibited metallic behavior down to 2 K, with resistance decreasing as temperature decreases. The room-temperature sheet resistance systematically
decreased with increasing $\mathrm{Ar}^{+}$ ion irradiation time [see Fig. ~\ref{fig1}(d)], a trend we attribute to the higher concentration of oxygen vacancies. A longer $\mathrm{Ar}^+$ ion irradiation time introduces a greater number of oxygen vacancies in the KTO, leading to a higher electron concentration and reduction in sheet resistance. The observed metallicity confirms the successful creation of a 2DEG.

Atomic force microscopy (AFM) was used to analyze the surface morphology of all samples [Fig.~\ref{fig2}(a, b)]. The non-irradiated KTO substrate had an atomically flat surface with a root-mean-square (RMS) roughness of $\sim 0.27 \mathrm{~nm}$ over a $2 \mu \mathrm{m} \times 2 \mu \mathrm{m}$ area. While the roughness remained below $1~\text{nm}$ for all the samples, it was found to increase slightly to $0.57
~\text{nm}$ as the irradiation time extended to $20$ minutes (Fig.~\ref{fig2}(b)).

Figure ~\ref{fig2}(c,d) displays the XPS analysis of the Ta 4f core levels in a KTO single-crystal substrate, both before and after $\mathrm{Ar}^+$ ion irradiation. The spectrum from the unirradiated KTO is successfully fitted with a single $\mathrm{Ta}^{5+}$ doublet ( $4 \mathrm{f}_{7 / 2}$ at $\sim 26~ \mathrm{eV}$ and $4 \mathrm{f}_{5 / 2}$ at 28~$\mathrm{eV}$ ), matching literature values and confirming a perfect initial stoichiometry with no oxygen vacancies ~\cite{VicenteArche2021}. In contrast, the spectrum for the KTO substrate irradiated with $\mathrm{Ar}^+$ for 20 minutes exhibits a low-energy shoulder [see Fig.~\ref{fig2}(d)]. Deconvolution of this shoulder-like feature reveals the presence of reduced Ta species: $\mathrm{Ta}^{4+}$ (contributing $14.8 \%$ of the total Ta area) and $\mathrm{Ta}^{2+}$ (contributing 8.5\% area). This valence state reduction, from $\mathrm{Ta}^{5+}$ to $\mathrm{Ta}^{4+}$ and $\mathrm{Ta}^{2+}$, is a direct spectroscopic signature of oxygen vacancy formation and the resultant electron doping into the Ta $5d$ bands. This electron doping mechanism is consistent with previous XPS studies on KTO substrates\cite{Mallik2022,Martnez2023}.

\subsection{Ferromagnetic Resonance (FMR) Measurements}

\begin{figure*}[htb]
    \centering
    \includegraphics[width=0.8\linewidth]{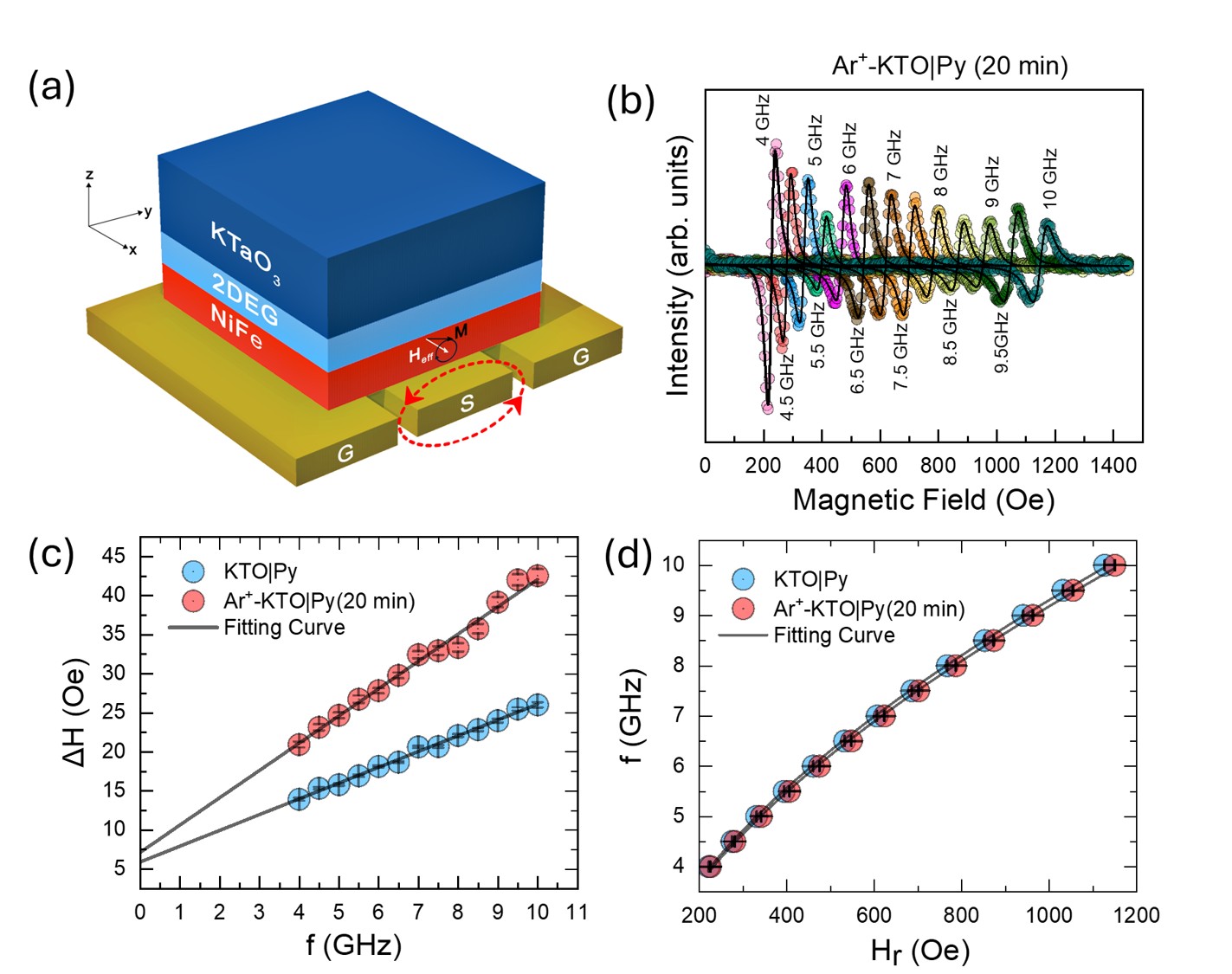}
    \caption{(a) Experimental schematic depicting the configuration used for the ferromagnetic resonance (FMR) measurements. (b) Room temperature FMR derivative absorption spectra $(d I(H) / d H$, where $I$ is the microwave absorption intensity) recorded for microwave frequencies ranging from 4 to 10 GHz (in 0.5 GHz steps). The solid lines represent the Lorentzian best fits to the experimental data. (c) Frequency dependence of the FMR linewidth $(\Delta H)$ for the reference KTO/Py bilayer and the 20min $\mathrm{Ar}^{+}$-irradiated $\mathrm{Ar}^{+}-\mathrm{KTO} / \mathrm{Py}$ bilayer. Solid lines are fits to Equation~\ref{eq4}. (d) FMR resonance frequency ( $f$ ) plotted against the resonance field ( $H_{r}$ ) for both the reference and $\mathrm{Ar}^{+}-\mathrm{KTO} / \mathrm{Py}(20 \mathrm{~min})$ samples. The solid line represents the fit using the Kittel equation (Eq.~\ref{eq3}).}
    \label{fig3}
\end{figure*}

To investigate the spin-to-charge conversion properties of the bilayers, we performed room-temperature FMR measurements on the $\mathrm{Ar}^+-\mathrm{KTO} / \mathrm{Py}$ samples. The experimental setup consisted of a custom-built broadband spectrometer equipped with a coplanar waveguide, as illustrated in Fig.~\ref {fig3}(a)\cite{Bansal2018}. We carried out the FMR spectroscopy measurements in the radio frequency (RF) range of $4-10$ GHz at a power of +8 dBm. A representative differential absorption spectrum for an $\mathrm{Ar}^+-\mathrm{KTO} / \mathrm{Py}$ bilayer sample, irradiated for 20 minutes, is shown in Fig.~\ref {fig3}(b). This spectrum was recorded as a function of the external DC magnetic field $(H)$ at various microwave excitation frequencies $(f)$. The key FMR parameters-the resonance magnetic field ( $H_{r}$ ) and the linewidth ( $\Delta H$ )-were extracted by fitting the FMR spectra with a differential Lorentzian equation that included both symmetric and asymmetric components given by \cite{Kumar2019}

\begin{widetext}
\begin{equation}
\cfrac{d I}{d H}~\propto -2k_{s} \cdot \cfrac{\left( \Delta H \right)\left(H-H_{r}\right)}{\left[\left(H-H_{r}\right)^{2}+\left(\Delta H\right)^{2}\right]^{2}}+
k_{a s} \cdot \cfrac{\left(\Delta H\right)^{2}-\left(H-H_{r}\right)^{2}}{\left[\left(H-H_{r}\right)^{2}+\left(\Delta H\right)^{2}\right]^{2}}+
  H \times slope +\text { offset}. \label{eq2}
\end{equation}
\end{widetext}

Here, $\cfrac{d I}{d H}$ is the differential absorption intensity, $H$ is the applied DC magnetic field, and $k_{s}$ and $k_{a s}$ are the coefficients for the symmetric and asymmetric components of the line shape, respectively. To confirm that the spin-pumping measurements were performed within the linear excitation regime, we conducted microwave power-dependent FMR measurements over a range of +4 dBm to +10 dBm. These results and the corresponding analysis are detailed in Section S3 of the Supplementary Material.

Figure ~\ref{fig3}(c and d) shows the plot of linewidth ($\Delta H$) and the resonance field ($H_{r}$)  versus resonance frequency ($f$), respectively. The frequency dependence of the resonance field ($H_{r}$) follows a quadratic behavior corresponding to Kittel equation given as\cite{Kittel1948}:

\begin{equation}
f=\frac{\gamma}{2 \pi} \sqrt{\left(H_{r}+H_{K}\right)\left(H_{r}+H_{K}+4 \pi M_{e f f}\right)}  \label{eq3}
\end{equation}

\begin{figure*}[htb]
    \centering
    \includegraphics[width=0.9\linewidth]{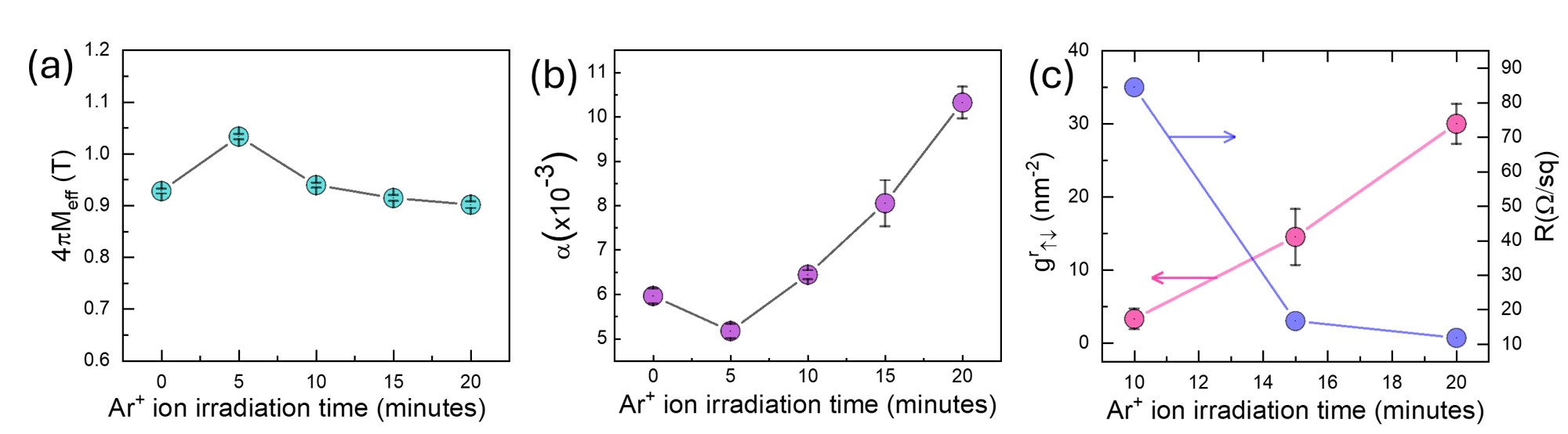}
    \caption{(a) Effective Magnetization ( $M_{e f f}$ ) as a function of $\mathrm{Ar}^+$ ion irradiation duration. The $M_{e f f}$ values were derived by applying a fit of Equation~\ref{eq3} to the FMR resonance frequency ( $f$ ) as a function of the resonance field $\left(H_{r}\right)$. (b) Effective Gilbert damping constant $(\alpha)$ versus $\mathrm{Ar}^{+}$ion irradiation time. The damping constant was calculated by fitting the frequency dependence of the FMR linewidth ( $\Delta H$ vs. $f$ ) using Equation~\ref{eq4}. (c) Spin-mixing conductance $\left(g_{\uparrow \downarrow}^r\right)$ and room-temperature sheet resistance as a function of $Ar^{+}$ ion irradiation time for KTO samples. The left axis shows $\left(g_{\uparrow \downarrow}^r\right)$ values calculated via Eq.~\ref{eq5}, while the right axis displays the corresponding sheet resistance.}
    \label{fig4}
\end{figure*}

Where $H_{K}$ is the anisotropy field, $4 \pi M_{\text {eff }}$ is the effective saturation magnetization, and $\gamma=\frac{g^* \mu_{B}}{\hbar}= 1.85 \times 10^{2} ~G H z T^{-1}$ is the gyromagnetic ratio for the Py. Here $g^{*}$ is the effective electron g-factor (taken to be 2.11 for Py), $\hbar$ is the reduced Planck constant, and $\mu_{B}$ is the Bohr magneton. Based on the fit of the data in Fig. ~\ref{fig3}(d) to the Kittel equation (Eq.~\ref{eq3}), we determined the effective saturation magnetization $4 \pi M_{e f f} \sim 0.9 ~\mathrm{T}$ for the 20 -minute $\mathrm{Ar}^{+}$irradiated sample. The values for $4 \pi M_{e f f}$ were found to be similar across all samples, as shown in Fig.~\ref {fig4}(a). The fitted values for the magnetocrystalline anisotropy field ( $H_{K}$ ) were found to be in the range of $20-30~ \mathrm{Oe}$ for all the $\mathrm{Ar}^{+}$-irradiated samples.

The FMR linewidth ($\Delta H$) as a function of microwave frequency ($f$) is plotted in Fig.~\ref {fig3}(c). The linear relationship observed confirms predominantly Gilbert-type damping, and extrinsic mechanisms such as the two-magnon scattering mechanism have a negligible contribution to the relaxation in our samples~\cite{Arias1999}. The effective damping constant ($\alpha_{Ar^{+}-KTO/Py}$) can be determined from the slope using the following expression~\cite{Rossing1963,Heinrich1985}.

\begin{equation}
\Delta H=\frac{2 \pi \alpha_{A r^{+}-K T O / P y} f}{\gamma}+\Delta H_{0}. \label{eq4}
\end{equation}

Where $\gamma$ is the gyromagnetic ratio of the electron in Py. Here, the first term represents the frequency-dependent contribution to magnetization relaxation, which arises from both the intrinsic Gilbert damping and the spin-pumping mechanism. The second term, $\Delta H_{0}$, is the frequency-independent contribution, known as inhomogeneous linewidth broadening. This term is a result of structural inhomogeneities or defects within the Py film. The inhomogeneous linewidth broadening ($\Delta H_{0}$) is obtained from the y-intercept of the fit at $f=0$. The extracted $\Delta H_0$ values ranged from 3 to 7 Oe for all samples. Given that $\Delta H_0$ is typically associated with surface roughness and structural defects, the low magnitude of $\Delta H_0$ is consistent with the high degree of surface smoothness observed in our samples. This indicates that surface-related inhomogeneities play a negligible role in the total linewidth broadening.

We calculated the effective Gilbert damping constant, $\alpha_{A r^{+}-K T O / P y}$, for all the $\mathrm{Ar}^{+}-\mathrm{KTO} / \mathrm{Py}$ samples from the slope of the linear fit between the FMR linewidth ($\Delta H$) and frequency ($f$), as shown in Fig. ~\ref{fig3}(c). The two linear fits yielded $\alpha_{K T O / P y}=5.98 \times 10^{-3}$ for the virgin KTO/Py reference sample and $\alpha_{A r^{+}-K T O / P y}=1.03 \times 10^{-2}$ for the 20 minutes irradiated $\mathrm{Ar}^{+}-\mathrm{KTO} / \mathrm{Py}$ sample. The enhancement in Gilbert damping constant for the irradiated sample compared to the pristine reference is consistent with the transfer of spin angular momentum (spin pumping) from the Py layer to the KTO 2DEG. As demonstrated in Supplementary Section S4, the contributions from radiative and eddy-current damping are several orders of magnitude smaller than the measured Gilbert damping constant values. Consequently, these extrinsic factors were excluded to focus the analysis on the dominant intrinsic and spin-pumping-induced damping channels.

The observed $\alpha_{A r^{+}-K T O / P y}$ values are comparable to those reported in similar systems~\cite{Al-Tawhid2025}. For instance, the effective damping constant for $\mathrm{Ar}^{+}$-irradiated STO/Py 2DEG has been reported as $\alpha_{A r^{+}-S T O / P y}=$ 0.01, while for an epitaxial STO/LAO/Py 2DEG, the value is $\alpha_{S T O / L A O / P y}= 0.0011$~\cite{Yang2019}. The Gilbert damping coefficient $\alpha_{A r^{+}-K T O / P y}$ was further estimated for $\mathrm{Ar}^{+}-\mathrm{KTO} / \mathrm{Py}$ thin films with varying $\mathrm{Ar}^{+}$ion irradiation times, and the results are summarized in Fig. ~\ref{fig4}(b). Detailed fitting procedures for all samples can be found in the supplementary material (Section S2). Interestingly, as the $\mathrm{Ar}^{+}$ion irradiation time increased, the damping coefficient $\alpha_{A r^{+}-K T O / P y}$ initially decreased relative to the unirradiated KTO/Py sample ($\alpha_{K T O / P y}$) before increasing significantly with longer irradiation times. The initial decrease in $\alpha_{A r^{+}-K T O / P y}$ observed at 5 minutes of $\mathrm{Ar}^{+}$ion irradiation may be related to the formation of an amorphous layer on the KTO surface in the early stages of the irradiation process~\cite{Wadehra2017}. It is important to note that the 2DEG is formed beneath this amorphous layer as the irradiation time is further increased beyond 5 minutes.

To evaluate the spin-pumping efficiency for the KTO/Py bilayers, we calculated the real part of the effective spin-mixing conductance $\left(g_{\uparrow \downarrow}^r\right)$, using the following expression\cite{Tserkovnyak2005}:

\begin{equation}
g_{\uparrow \downarrow}^r=\frac{4 \pi M_{s} t_{P y}}{g \mu_{B}}\left(\alpha_{A r^{+}-K T O \mid P y}-\alpha_{K T O \mid P y}\right) \label{eq5}
\end{equation}

Where $g$ is the Landé g-factor, $\mu_{B}$ is the Bohr magneton, $M_{s}$ is the saturation magnetization, and $t_{Py}$ is the thickness of the Py film. We found the spin-mixing conductance for the 10-minute $\mathrm{Ar}^{+}$-irradiated sample to be $g_{\uparrow \downarrow}^r=3.31\pm 1.42 \mathrm{~nm}^{-2}$. This value steadily increased, reaching a maximum value of $g_{\uparrow \downarrow}^r=30\pm 2.73 \mathrm{~nm}^{-2}$ for the sample irradiated for 20 minutes. The results for all samples with varying irradiation times are summarized in Fig.~\ref {fig4}(c). This value of spin-mixing conductance is comparable to that of other systems, such as the $\mathrm{Ar}^{+}$ion irradiated $\mathrm{Ar}^{+}$-STO/Py bilayer $\left(g_{\uparrow \downarrow}^r=29 \mathrm{~nm}^{-2}\right)$ ~\cite{Zhang2016}, AlO$_x$/STO/Py bilayer $\left(g_{\uparrow \downarrow}^r=2.2 \mathrm{~nm}^{-2}\right)$ \cite{Vaz2019}, and the epitaxial STO/LAO/Py system $\left(g_{\uparrow \downarrow}^r=13.3 \mathrm{~nm}^{-2}\right)$ \cite{Lesne2016}. It is also in the same order of magnitude as values reported for metallic systems like $\mathrm{Py} / \mathrm{Pt}\left(g_{\uparrow \downarrow}^r=8.6 \mathrm{~nm}^{-2}\right)$\cite{You2021} and $\mathrm{Py} / \mathrm{Ta}\left(g_{\uparrow \downarrow}^r=\right. 1.4 \mathrm{~nm}^{-2}$ ) \cite{Akash-prb-2018}, as well as other 2DEG systems like $\mathrm{Ag} / \mathrm{Bi}\left(g_{\uparrow \downarrow}^r=19 \mathrm{~nm}^{-2}\right)$ \cite{Snchez2013}.

The systematic increase in spin-mixing conductance $g_{\uparrow \downarrow}^r$ for $\mathrm{Ar}^{+}$irradiation times beyond 10 minutes is consistent with an increase in oxygen vacancy concentration, which enhances the conductivity of the 2DEG, as previously shown in Fig.~\ref {fig1}(d).

\section{Conclusions}
In summary, we successfully created a 2DEG on the KTO (001) surface using $\mathrm{Ar}^{+}$ ion irradiation. This process generates oxygen vacancies in the near-surface region, leading to electron doping as confirmed by XPS analysis. The metallic behavior of the resulting 2DEG layer was verified through temperature-dependent resistivity measurements. We performed the comprehensive spin-pumping experiments on this system, confirming successful spin current injection into the 2DEG by observing a significant enhancement of magnetic damping in the $\mathrm{Ar}^{+}$-irradiated KTO/Py bilayer compared to a non-irradiated control sample. Crucially, we showed that the spin-mixing conductance ($g_{\uparrow \downarrow}^r$), a key metric for spin current injection efficiency, can be substantially increased by controlling the $\mathrm{Ar}^{+}$irradiation duration. This improvement is a result of the increased oxygen vacancy concentration, which boosts 2DEG conductivity. The ability to control and boost the spin-mixing conductance through a simple, scalable technique like $\mathrm{Ar}^{+}$irradiation is a major step forward, offering a guideline for optimizing KTO-based spintronic systems. Our findings suggest $\mathrm{Ar}^{+}$-irradiated KTO 2DEGs as a promising platform for next-generation oxide spintronics\cite{Huihui2025,Trier2021}.

\section*{Data availability}
The data that support the findings of this study are available from the corresponding author upon reasonable request.

\section*{Author contributions}
\textbf{Yasar K. Arafath :} Data curation, Formal analysis, Investigation, Methodology, Validation, Writing – review \& editing.
\textbf{Vaisali Yadav:} Data curation, Methodology, Investigation, Validation, Formal analysis, Writing – review \& editing.
\textbf{Nidhi Kandwal:} Data curation, Methodology, Investigation, Validation, Formal analysis, Writing – review \& editing.
\textbf{P.N. Santhosh:} Funding acquisition, Resources, Supervision, Validation, Writing – review \& editing.
\textbf{Pranaba Kishore Muduli:} Formal analysis, Funding acquisition, Investigation, Methodology, Resources, Supervision, Validation, Writing – original draft, Writing – review \& editing.
\textbf{Prasanta Kumar Muduli:} Conceptualization, Formal analysis, Funding acquisition, Investigation, Methodology, Project administration, Resources, Supervision, Validation,  Writing – original draft, Writing – review \& editing.

\section*{Conflicts of interest}
There are no conflicts to declare.
\section*{Acknowledgements}
PKM acknowledges funding from IIT Madras (Grants No. IP21221798PHNFSC008989 and No. RF21221392PHNFIG008989) and Science and Engineering Research Board (SERB), India, Grant No. SRG/2022/000438.

YKA acknowledges funding from the Prime Minister Research Fellowship (PMRF) Scheme grant No.SB23242097PHPMRF008140.

The authors acknowledge the Institute of Eminence (IoE) initiative scheme by the Indian Institute of Technology, Madras, for the financial and logistical support of the FORG.

The authors acknowledge the DST-FIST funding (Project No. SR/FST/PSII-038/2016) for the PPMS facility in the Department of Physics, IIT Madras, India.

\bibliography{reference}

@article{Hwang2012,
   author = {H. Y. Hwang and Y. Iwasa and M. Kawasaki and B. Keimer and N. Nagaosa and Y. Tokura},
   doi = {10.1038/nmat3223},
   issn = {1476-1122},
   issue = {2},
   journal = {Nature Materials},
   month = {2},
   pages = {103-113},
   title = {Emergent phenomena at oxide interfaces},
   volume = {11},
   year = {2012}
}

@article{Chen2023,
   author = {Shiwei Chen and Chuantong Ren and Shiheng Liang},
   doi = {10.34133/adi.0024},
   issn = {2767-9713},
   journal = {Advanced Devices \& Instrumentation},
   month = {1},
   title = {Spintronics Phenomena of Two-Dimensional Electron Gas at Oxide Interfaces},
   volume = {4},
   year = {2023}
}

@article{Ohtomo2004,
   author = {A. Ohtomo and H. Y. Hwang},
   doi = {10.1038/nature02308},
   issn = {0028-0836},
   issue = {6973},
   journal = {Nature},
   month = {1},
   pages = {423-426},
   title = {A high-mobility electron gas at the {LaAlO$_3$}/{SrTiO$_3$} heterointerface},
   volume = {427},
   year = {2004}
}

@article{Varignon2018,
   author = {J. Varignon and L. Vila and A. Barthélémy and M. Bibes},
   doi = {10.1038/s41567-018-0112-1},
   issn = {1745-2473},
   issue = {4},
   journal = {Nature Physics},
   month = {4},
   pages = {322-325},
   title = {A new spin for oxide interfaces},
   volume = {14},
   year = {2018}
}

@article{Dikin2011,
   author = {D. A. Dikin and M. Mehta and C. W. Bark and C. M. Folkman and C. B. Eom and V. Chandrasekhar},
   doi = {10.1103/PhysRevLett.107.056802},
   issn = {0031-9007},
   issue = {5},
   journal = {Physical Review Letters},
   month = {7},
   pages = {056802},
   title = {Coexistence of Superconductivity and Ferromagnetism in Two Dimensions},
   volume = {107},
   year = {2011}
}

@article{Li2011,
   author = {Lu Li and C. Richter and J. Mannhart and R. C. Ashoori},
   doi = {10.1038/nphys2080},
   issn = {1745-2473},
   issue = {10},
   journal = {Nature Physics},
   month = {10},
   pages = {762-766},
   title = {Coexistence of magnetic order and two-dimensional superconductivity at LaAlO3/SrTiO3 interfaces},
   volume = {7},
   year = {2011}
}

@article{Reyren2007,
   author = {N. Reyren and S. Thiel and A. D. Caviglia and L. Fitting Kourkoutis and G. Hammerl and C. Richter and C. W. Schneider and T. Kopp and A.-S. Rüetschi and D. Jaccard and M. Gabay and D. A. Muller and J.-M. Triscone and J. Mannhart},
   doi = {10.1126/science.1146006},
   issn = {0036-8075},
   issue = {5842},
   journal = {Science},
   month = {8},
   pages = {1196-1199},
   title = {Superconducting Interfaces Between Insulating Oxides},
   volume = {317},
   year = {2007}
}

@article{Brhin2023,
   author = {Julien Bréhin and Yu Chen and Maria D’Antuono and Sara Varotto and Daniela Stornaiuolo and Cinthia Piamonteze and Julien Varignon and Marco Salluzzo and Manuel Bibes},
   doi = {10.1038/s41567-023-01983-y},
   issn = {1745-2473},
   issue = {6},
   journal = {Nature Physics},
   month = {6},
   pages = {823-829},
   title = {Coexistence and coupling of ferroelectricity and magnetism in an oxide two-dimensional electron gas},
   volume = {19},
   year = {2023}
}

@article{Rubi2020,
   author = {Km Rubi and Julien Gosteau and Raphaël Serra and Kun Han and Shengwei Zeng and Zhen Huang and Benedicte Warot-Fonrose and Rémi Arras and Etienne Snoeck and Ariando and Michel Goiran and Walter Escoffier},
   doi = {10.1038/s41535-020-0210-z},
   issn = {2397-4648},
   issue = {1},
   journal = {npj Quantum Materials},
   month = {1},
   pages = {9},
   title = {Aperiodic quantum oscillations in the two-dimensional electron gas at the {LaAlO$_3$}/{SrTiO$_3$} interface},
   volume = {5},
   year = {2020}
}

@article{Matsubara2016,
   author = {Y. Matsubara and K. S. Takahashi and M. S. Bahramy and Y. Kozuka and D. Maryenko and J. Falson and A. Tsukazaki and Y. Tokura and M. Kawasaki},
   doi = {10.1038/ncomms11631},
   issn = {2041-1723},
   issue = {1},
   journal = {Nature Communications},
   month = {5},
   pages = {11631},
   title = {Observation of the quantum Hall effect in δ-doped {SrTiO$_3$}},
   volume = {7},
   year = {2016}
}

@article{Tebano2012,
author = {Tebano, Antonello and Fabbri, Emiliana and Pergolesi, Daniele and Balestrino, Giuseppe and Traversa, Enrico},
title = {Room-Temperature Giant Persistent Photoconductivity in {SrTiO$_3$}/{LaAlO$_3$} Heterostructures},
journal = {ACS Nano},
volume = {6},
number = {2},
pages = {1278-1283},
year = {2012},
doi = {10.1021/nn203991q},
URL = {https://doi.org/10.1021/nn203991q},
}

@article{Caviglia2008,
   author = {A. D. Caviglia and S. Gariglio and N. Reyren and D. Jaccard and T. Schneider and M. Gabay and S. Thiel and G. Hammerl and J. Mannhart and J.-M. Triscone},
   doi = {10.1038/nature07576},
   issn = {0028-0836},
   issue = {7222},
   journal = {Nature},
   month = {12},
   pages = {624-627},
   title = {Electric field control of the {LaAlO$_3$}/{SrTiO$_3$} interface ground state},
   volume = {456},
   year = {2008}
}

@article{Lesne2016,
   author = {E. Lesne and Yu Fu and S. Oyarzun and J. C. Rojas-Sánchez and D. C. Vaz and H. Naganuma and G. Sicoli and J.-P. Attané and M. Jamet and E. Jacquet and J.-M. George and A. Barthélémy and H. Jaffrès and A. Fert and M. Bibes and L. Vila},
   doi = {10.1038/nmat4726},
   issn = {1476-1122},
   issue = {12},
   journal = {Nature Materials},
   month = {12},
   pages = {1261-1266},
   title = {Highly efficient and tunable spin-to-charge conversion through Rashba coupling at oxide interfaces},
   volume = {15},
   year = {2016}
}

@article{Herranz2007,
   author = {G. Herranz and M. Basletić and M. Bibes and C. Carrétéro and E. Tafra and E. Jacquet and K. Bouzehouane and C. Deranlot and A. Hamzić and J.-M. Broto and A. Barthélémy and A. Fert},
   doi = {10.1103/PhysRevLett.98.216803},
   issn = {0031-9007},
   issue = {21},
   journal = {Physical Review Letters},
   month = {5},
   pages = {216803},
   title = {High Mobility in {LaAlO$_3$}/{SrTiO$_3$} Heterostructures: Origin, Dimensionality, and Perspectives},
   volume = {98},
   year = {2007}
}

@article{Gallagher2015,
   author = {Patrick Gallagher and Menyoung Lee and Trevor A. Petach and Sam W. Stanwyck and James R. Williams and Kenji Watanabe and Takashi Taniguchi and David Goldhaber-Gordon},
   doi = {10.1038/ncomms7437},
   issn = {2041-1723},
   issue = {1},
   journal = {Nature Communications},
   month = {3},
   pages = {6437},
   title = {A high-mobility electronic system at an electrolyte-gated oxide surface},
   volume = {6},
   year = {2015}
}

@article{Zhang2017,
   author = {Hui Zhang and Hongrui Zhang and Xi Yan and Xuejing Zhang and Qinghua Zhang and Jing Zhang and Furong Han and Lin Gu and Banggui Liu and Yuansha Chen and Baogen Shen and Jirong Sun},
   doi = {10.1021/acsami.7b12814},
   issn = {1944-8244},
   issue = {41},
   journal = {ACS Applied Materials \& Interfaces},
   month = {10},
   pages = {36456-36461},
   title = {Highly Mobile Two-Dimensional Electron Gases with a Strong Gating Effect at the Amorphous {LaAlO$_3$}/{KTaO$_3$} Interface},
   volume = {9},
   year = {2017}
}

@article{Rdel2016,
   author = {Tobias Chris Rödel and Franck Fortuna and Shamashis Sengupta and Emmanouil Frantzeskakis and Patrick Le Fèvre and François Bertran and Bernard Mercey and Sylvia Matzen and Guillaume Agnus and Thomas Maroutian and Philippe Lecoeur and Andrés Felipe Santander‐Syro},
   doi = {10.1002/adma.201505021},
   issn = {0935-9648},
   issue = {10},
   journal = {Advanced Materials},
   month = {3},
   pages = {1976-1980},
   title = {Universal Fabrication of 2D Electron Systems in Functional Oxides},
   volume = {28},
   year = {2016}
}

@article{Vicente-Arche2021,
   author = {Luis M. Vicente-Arche and Srijani Mallik and Maxen Cosset-Cheneau and Paul Noël and Diogo C. Vaz and Felix Trier and Tanay A. Gosavi and Chia-Ching Lin and Dmitri E. Nikonov and Ian A. Young and Anke Sander and Agnès Barthélémy and Jean-Philippe Attané and Laurent Vila and Manuel Bibes},
   doi = {10.1103/PhysRevMaterials.5.064005},
   issn = {2475-9953},
   issue = {6},
   journal = {Physical Review Materials},
   month = {6},
   pages = {064005},
   title = {Metal/{SrTiO$_3$}
 two-dimensional electron gases for spin-to-charge conversion},
   volume = {5},
   year = {2021}
}

@article{Meevasana2011,
   author = {W. Meevasana and P. D. C. King and R. H. He and S-K. Mo and M. Hashimoto and A. Tamai and P. Songsiriritthigul and F. Baumberger and Z-X. Shen},
   doi = {10.1038/nmat2943},
   issn = {1476-1122},
   issue = {2},
   journal = {Nature Materials},
   month = {2},
   pages = {114-118},
   title = {Creation and control of a two-dimensional electron liquid at the bare {SrTiO$_3$} surface},
   volume = {10},
   year = {2011}
}

@article{Reagor2005,
   author = {David W. Reagor and Vladimir Y. Butko},
   doi = {10.1038/nmat1402},
   issn = {1476-1122},
   issue = {8},
   journal = {Nature Materials},
   month = {8},
   pages = {593-596},
   title = {Highly conductive nanolayers on strontium titanate produced by preferential ion-beam etching},
   volume = {4},
   year = {2005}
}

@article{Chang2015,
   author = {Jung-Won Chang and Joon Sung Lee and Tae Ho Lee and Jinhee Kim and Yong-Joo Doh},
   doi = {10.7567/APEX.8.055701},
   issn = {1882-0778},
   issue = {5},
   journal = {Applied Physics Express},
   month = {5},
   pages = {055701},
   title = {Controlled formation of high-mobility shallow electron gases in {SrTiO$_3$} single crystal},
   volume = {8},
   year = {2015}
}

@article{ QiuruWang2016,
   author = {Qiuru Wang and Wanli Zhang and Wenxu Zhang and Huizhong Zeng},
   doi = {10.1016/j.apsusc.2015.12.219},
   issn = {01694332},
   journal = {Applied Surface Science},
   month = {3},
   pages = {84-87},
   title = {In-situ monitor of insulator to metal transition in SrTiO3 by Ar+ irradiation},
   volume = {365},
   year = {2016}
}

@article{Santander-Syro2011-mw,
   author = {A. F. Santander-Syro and O. Copie and T. Kondo and F. Fortuna and S. Pailhès and R. Weht and X. G. Qiu and F. Bertran and A. Nicolaou and A. Taleb-Ibrahimi and P. Le Fèvre and G. Herranz and M. Bibes and N. Reyren and Y. Apertet and P. Lecoeur and A. Barthélémy and M. J. Rozenberg},
   doi = {10.1038/nature09720},
   issn = {0028-0836},
   issue = {7329},
   journal = {Nature},
   month = {1},
   pages = {189-193},
   title = {Two-dimensional electron gas with universal subbands at the surface of SrTiO3},
   volume = {469},
   year = {2011}
}

@article{Gupta2022,
   author = {Anshu Gupta and Harsha Silotia and Anamika Kumari and Manish Dumen and Saveena Goyal and Ruchi Tomar and Neha Wadehra and Pushan Ayyub and Suvankar Chakraverty},
   doi = {10.1002/adma.202106481},
   issn = {0935-9648},
   issue = {9},
   journal = {Advanced Materials},
   month = {3},
   title = {{KTaO$_3$} —The New Kid on the Spintronics Block},
   volume = {34},
   year = {2022}
}

@article{Zhang2023,
   author = {Hui Zhang and Zengtai Zhu and Yungu Zhu and Xiaobing Chen and Qisheng Jiang and Jinwu Wei and Chenbo Zhao and Jine Zhang and Furong Han and Huaiwen Yang and Dapeng Zhu and Hao Wu and Yuansha Chen and Fengxia Hu and Baogen Shen and Weisheng Zhao and Jing Zhang and Guoqiang Yu and Jirong Sun},
   doi = {10.1103/PhysRevApplied.19.034045},
   issn = {2331-7019},
   issue = {3},
   journal = {Physical Review Applied},
   month = {3},
   pages = {034045},
   title = {Fermi-Level-Dependent Charge-to-Spin Conversion of the Two-Dimensional Electron Gas at the {$\gamma$-Al$_2$O$_3$}/{KTaO$_3$} Interface},
   volume = {19},
   year = {2023}
}

@article{Wadehra2020,
   author = {Neha Wadehra and Ruchi Tomar and Rahul Mahavir Varma and R. K. Gopal and Yogesh Singh and Sushanta Dattagupta and S. Chakraverty},
   doi = {10.1038/s41467-020-14689-z},
   issn = {2041-1723},
   issue = {1},
   journal = {Nature Communications},
   month = {2},
   pages = {874},
   title = {Planar Hall effect and anisotropic magnetoresistance in polar-polar interface of {LaVO$_3$}-{KTaO$_3$} with strong spin-orbit coupling},
   volume = {11},
   year = {2020}
}

@article{Qi2024,
   author = {Shaojin Qi and Jiexing Liang and Guimei Shi and Yulin Gan and Yuansha Chen and Yunzhong Chen and Jirong Sun},
   doi = {10.1021/acsaelm.4c01618},
   issn = {2637-6113},
   issue = {11},
   journal = {ACS Applied Electronic Materials},
   month = {11},
   pages = {8404-8412},
   title = {Creation of Two-Dimensional Electron Gas at the Heterointerface of {CaZrO$_3$}/{KTaO$_3$} with Tunable Rashba Spin–Orbit Coupling},
   volume = {6},
   year = {2024}
}

@article{Zou2015,
   author = {K. Zou and Sohrab Ismail-Beigi and Kim Kisslinger and Xuan Shen and Dong Su and F. J. Walker and C. H. Ahn},
   doi = {10.1063/1.4914310},
   issn = {2166-532X},
   issue = {3},
   journal = {APL Materials},
   month = {3},
   title = {{LaTiO$_3$}/{KTaO$_3$}interfaces: A new two-dimensional electron gas system},
   volume = {3},
   year = {2015}
}

@article{Hua2022,
   author = {Xiangyu Hua and Fanbao Meng and Zongyao Huang and Zhaohang Li and Shuai Wang and Binghui Ge and Ziji Xiang and Xianhui Chen},
   doi = {10.1038/s41535-022-00506-x},
   issn = {2397-4648},
   issue = {1},
   journal = {npj Quantum Materials},
   month = {9},
   pages = {97},
   title = {Tunable two-dimensional superconductivity and spin-orbit coupling at the {EuO}/{KTaO$_3$(110)} interface},
   volume = {7},
   year = {2022}
}

@article{Liu2021,
   author = {Changjiang Liu and Xi Yan and Dafei Jin and Yang Ma and Haw-Wen Hsiao and Yulin Lin and Terence M. Bretz-Sullivan and Xianjing Zhou and John Pearson and Brandon Fisher and J. Samuel Jiang and Wei Han and Jian-Min Zuo and Jianguo Wen and Dillon D. Fong and Jirong Sun and Hua Zhou and Anand Bhattacharya},
   doi = {10.1126/science.aba5511},
   issn = {0036-8075},
   issue = {6530},
   journal = {Science},
   month = {2},
   pages = {716-721},
   title = {Two-dimensional superconductivity and anisotropic transport at {KTaO$_3$(111)} interfaces},
   volume = {371},
   year = {2021}
}

@article{Nakamura2009,
   author = {H. Nakamura and T. Kimura},
   doi = {10.1103/PhysRevB.80.121308},
   issn = {1098-0121},
   issue = {12},
   journal = {Physical Review B},
   month = {9},
   pages = {121308},
   title = {Electric field tuning of spin-orbit coupling in {KTaO$_3$}
 field-effect transistors},
   volume = {80},
   year = {2009}
}

@article{Ueno2011,
   author = {K. Ueno and S. Nakamura and H. Shimotani and H. T. Yuan and N. Kimura and T. Nojima and H. Aoki and Y. Iwasa and M. Kawasaki},
   doi = {10.1038/nnano.2011.78},
   issn = {1748-3387},
   issue = {7},
   journal = {Nature Nanotechnology},
   month = {7},
   pages = {408-412},
   title = {Discovery of superconductivity in {KTaO$_3$} by electrostatic carrier doping},
   volume = {6},
   year = {2011}
}

@article{Harashima2013,
  title = {Coexistence of two-dimensional and three-dimensional Shubnikov--de Haas oscillations in Ar${}^{+}$-irradiated KTaO${}_{3}$},
  author = {Harashima, S. and Bell, C. and Kim, M. and Yajima, T. and Hikita, Y. and Hwang, H. Y.},
  journal = {Phys. Rev. B},
  volume = {88},
  issue = {8},
  pages = {085102},
  numpages = {7},
  year = {2013},
  month = {Aug},
  publisher = {American Physical Society},
  doi = {10.1103/PhysRevB.88.085102},
  url = {https://link.aps.org/doi/10.1103/PhysRevB.88.085102}
}

@article{Han2024,
   author = {Yamin Han and Bin Lao and Xuan Zheng and Sheng Li and Run-Wei Li and Zhiming Wang},
   doi = {10.3389/fmats.2024.1444769},
   issn = {2296-8016},
   journal = {Frontiers in Materials},
   month = {8},
   title = {Transition metal oxides: a new frontier in spintronics driven by novel quantum states and efficient charge-spin interconversion},
   volume = {11},
   year = {2024}
}

@article{Zhang2019,
   author = {Hongrui Zhang and Yang Ma and Hui Zhang and Xiaobing Chen and Shuanhu Wang and Gang Li and Yu Yun and Xi Yan and Yuansha Chen and Fengxia Hu and Jianwang Cai and Baogen Shen and Wei Han and Jirong Sun},
   doi = {10.1021/acs.nanolett.8b04509},
   issn = {1530-6984},
   issue = {3},
   journal = {Nano Letters},
   month = {3},
   pages = {1605-1612},
   title = {Thermal Spin Injection and Inverse Edelstein Effect of the Two-Dimensional Electron Gas at {EuO}/{KTaO$_3$}
 Interfaces},
   volume = {19},
   year = {2019}
}

@article{VicenteArche2021,
   author = {Luis M. Vicente‐Arche and Julien Bréhin and Sara Varotto and Maxen Cosset‐Cheneau and Srijani Mallik and Raphaël Salazar and Paul Noël and Diogo C. Vaz and Felix Trier and Suvam Bhattacharya and Anke Sander and Patrick Le Fèvre and François Bertran and Guilhem Saiz and Gerbold Ménard and Nicolas Bergeal and Agnès Barthélémy and Hai Li and Chia‐Ching Lin and Dmitri E. Nikonov and Ian A. Young and Julien E. Rault and Laurent Vila and Jean‐Philippe Attané and Manuel Bibes},
   doi = {10.1002/adma.202102102},
   issn = {0935-9648},
   issue = {43},
   journal = {Advanced Materials},
   month = {10},
   title = {Spin–Charge Interconversion in {KTaO$_3$}
 2D Electron Gases},
   volume = {33},
   year = {2021}
}

@article{Al-Tawhid2025,
   author = {Athby H. Al-Tawhid and Rui Sun and Andrew H. Comstock and Divine P. Kumah and Dali Sun and Kaveh Ahadi},
   doi = {10.1063/5.0247001},
   issn = {0003-6951},
   issue = {9},
   journal = {Applied Physics Letters},
   month = {3},
   title = {Spin-to-charge conversion at {KTaO$_3$(111)} interfaces},
   volume = {126},
   year = {2025}
}

@article{Zheng2024,
   author = {Dongyao Zheng and Hui Zhang and Fengxia Hu and Baogen Shen and Jirong Sun and Weisheng Zhao},
   doi = {10.1088/1361-6528/ad0dca},
   issn = {0957-4484},
   issue = {9},
   journal = {Nanotechnology},
   month = {2},
   pages = {092001},
   title = {Spin-charge interconversion of two-dimensional electron gases at oxide interfaces},
   volume = {35},
   year = {2024}
}

@article{Saitoh2006,
   author = {E. Saitoh and M. Ueda and H. Miyajima and G. Tatara},
   doi = {10.1063/1.2199473},
   issn = {0003-6951},
   issue = {18},
   journal = {Applied Physics Letters},
   month = {5},
   title = {Conversion of spin current into charge current at room temperature: Inverse spin-Hall effect},
   volume = {88},
   year = {2006}
}

@article{Mosendz2010,
   author = {O. Mosendz and J. E. Pearson and F. Y. Fradin and G. E. W. Bauer and S. D. Bader and A. Hoffmann},
   doi = {10.1103/PhysRevLett.104.046601},
   issn = {0031-9007},
   issue = {4},
   journal = {Physical Review Letters},
   month = {1},
   pages = {046601},
   title = {Quantifying Spin Hall Angles from Spin Pumping: Experiments and Theory},
   volume = {104},
   year = {2010}
}

@article{Swindells2022,
   author = {C. Swindells and D. Atkinson},
   doi = {10.1063/5.0080267},
   issn = {0021-8979},
   issue = {17},
   journal = {Journal of Applied Physics},
   month = {5},
   title = {Interface enhanced precessional damping in spintronic multilayers: A perspective},
   volume = {131},
   year = {2022}
}

@article{Tserkovnyak2005,
   author = {Yaroslav Tserkovnyak and Arne Brataas and Gerrit E. W. Bauer and Bertrand I. Halperin},
   doi = {10.1103/RevModPhys.77.1375},
   issn = {0034-6861},
   issue = {4},
   journal = {Reviews of Modern Physics},
   month = {12},
   pages = {1375-1421},
   title = {Nonlocal magnetization dynamics in ferromagnetic heterostructures},
   volume = {77},
   year = {2005}
}

@article{Vaz2018,
   author = {Diogo Castro Vaz and Agnès Barthélémy and Manuel Bibes},
   doi = {10.7567/JJAP.57.0902A4},
   issn = {0021-4922},
   issue = {9},
   journal = {Japanese Journal of Applied Physics},
   month = {9},
   pages = {0902A4},
   title = {Oxide spin-orbitronics: New routes towards low-power electrical control of magnetization in oxide heterostructures},
   volume = {57},
   year = {2018}
}

@article{Bansal2018,
    author = {Bansal, Rajni and Chowdhury, Niru and Muduli, P. K.},
    title = {Proximity effect induced enhanced spin pumping in Py/Gd at room temperature},
    journal = {Applied Physics Letters},
    volume = {112},
    number = {26},
    pages = {262403},
    year = {2018},
    month = {06},
    issn = {0003-6951},
    doi = {10.1063/1.5033418},
    url = {https://doi.org/10.1063/1.5033418},
   
}

@article{Akash-prb-2018,
  title = {Large spin current generation by the spin Hall effect in mixed crystalline phase {Ta} thin films},
  author = {Kumar, Akash and Bansal, Rajni and Chaudhary, Sujeet and Muduli, Pranaba Kishor},
  journal = {Phys. Rev. B},
  volume = {98},
  issue = {10},
  pages = {104403},
  numpages = {6},
  year = {2018},
  month = {Sep},
  publisher = {American Physical Society},
  doi = {10.1103/PhysRevB.98.104403},
  url = {https://link.aps.org/doi/10.1103/PhysRevB.98.104403}
}

@article{Henrich1978,
  title = {Surface defects and the electronic structure of {SrTiO$_3$} surfaces},
  author = {Henrich, Victor E. and Dresselhaus, G. and Zeiger, H. J.},
  journal = {Phys. Rev. B},
  volume = {17},
  issue = {12},
  pages = {4908--4921},
  numpages = {0},
  year = {1978},
  month = {Jun},
  publisher = {American Physical Society},
  doi = {10.1103/PhysRevB.17.4908},
  url = {https://link.aps.org/doi/10.1103/PhysRevB.17.4908}
}

@ARTICLE{Wakabayashi2025,
	author = {Wakabayashi, Yuki K. and Krockenberger, Yoshiharu and Takiguchi, Kosuke and Yamamoto, Hideki and Taniyasu, Yoshitaka},
	title = {Role of ion milling angle in determining conducting and insulating states on {SrTiO$_3$} surfaces},
	year = {2025},
	journal = {Journal of Applied Physics},
	volume = {137},
	number = {9},
	doi = {10.1063/5.0245726},
	url = {https://www.scopus.com/inward/record.uri?eid=2-s2.0-86000283407&doi=10.1063%2f5.0245726&partnerID=40&md5=410b3ae131278ba4c2d803c869052e01},
	type = {Article},
	publication_stage = {Final},
	source = {Scopus},
	
}

@article{Gross2011,
   author = {Heiko Gross and Namrata Bansal and Yong-Seung Kim and Seongshik Oh},
   doi = {10.1063/1.3650254},
   issn = {0021-8979},
   issue = {7},
   journal = {Journal of Applied Physics},
   month = {10},
   title = {In situ study of emerging metallicity on ion-bombarded {SrTiO$_3$} surface},
   volume = {110},
   year = {2011}
}

@article{Psiuk2016,
   author = {B. Psiuk and J. Szade and K. Szot},
   doi = {10.1016/j.vacuum.2016.05.026},
   issn = {0042207X},
   journal = {Vacuum},
   month = {9},
   pages = {14-21},
   title = {{SrTiO$_3$} surface modification upon low energy {Ar$^+$} bombardment studied by XPS},
   volume = {131},
   year = {2016}
}

@article{Zhang2016,
    author = {Zhang, Wenxu and Wang, Qiuru and Peng, Bin and Zeng, Huizhong and Soh, Wee Tee and Ong, Chong Kim and Zhang, Wanli},
    title = {Spin galvanic effect at the conducting {SrTiO$_3$} surfaces},
    journal = {Applied Physics Letters},
    volume = {109},
    number = {26},
    pages = {262402},
    year = {2016},
    month = {12},
    issn = {0003-6951},
    doi = {10.1063/1.4973479},
    url = {https://doi.org/10.1063/1.4973479},
   
}

@article{Wadehra2017,
   author = {Neha Wadehra and Ruchi Tomar and Soumyadip Halder and Minaxi Sharma and Inderjit Singh and Nityasagar Jena and Bhanu Prakash and Abir De Sarkar and Chandan Bera and Ananth Venkatesan and S. Chakraverty},
   doi = {10.1103/PhysRevB.96.115423},
   issn = {2469-9950},
   issue = {11},
   journal = {Physical Review B},
   month = {9},
   pages = {115423},
   title = {Electronic structure modification of the {KTaO$_3$} single-crystal surface by {Ar$^+$} bombardment},
   volume = {96},
   year = {2017}
}

@article{DKumar2015,
  title = {Dynamics of photogenerated nonequilibrium electronic states in ${\mathrm{Ar}}^{+}$-ion-irradiated ${\mathrm{SrTiO}}_{3}$},
  author = {Kumar, Dushyant and Hossain, Z. and Budhani, R. C.},
  journal = {Phys. Rev. B},
  volume = {91},
  issue = {20},
  pages = {205117},
  numpages = {8},
  year = {2015},
  month = {May},
  publisher = {American Physical Society},
  doi = {10.1103/PhysRevB.91.205117},
  url = {https://link.aps.org/doi/10.1103/PhysRevB.91.205117}
}

@article{Schultz2007,
   author = {Moty Schultz and Lior Klein},
   doi = {10.1063/1.2795336},
   issn = {0003-6951},
   issue = {15},
   journal = {Applied Physics Letters},
   month = {10},
   title = {Relaxation of transport properties in electron-doped SrTiO3},
   volume = {91},
   year = {2007}
}

@article{Mallik2022,
   author = {S. Mallik and G. C. Ménard and G. Saïz and H. Witt and J. Lesueur and A. Gloter and L. Benfatto and M. Bibes and N. Bergeal},
   doi = {10.1038/s41467-022-32242-y},
   issn = {2041-1723},
   issue = {1},
   journal = {Nature Communications},
   month = {8},
   pages = {4625},
   title = {Superfluid stiffness of a {KTaO$_3$}
-based two-dimensional electron gas},
   volume = {13},
   year = {2022}
}

@article{Martnez2023,
   author = {Emanuel A. Martínez and Ji Dai and Massimo Tallarida and Norbert M. Nemes and Flavio Y. Bruno},
   doi = {10.1002/aelm.202300267},
   issn = {2199-160X},
   issue = {10},
   journal = {Advanced Electronic Materials},
   month = {10},
   title = {Anisotropic Electronic Structure of the 2D Electron Gas at the {AlO$_x$}/{KTaO$_3$(110)} Interface},
   volume = {9},
   year = {2023}
}

@article{Kumar2019,
   author = {Akash Kumar and Nidhi Pandey and Dileep Kumar and Mukul Gupta and Sujeet Chaudhary and Pranaba Kishor Muduli},
   doi = {10.1016/j.physb.2019.06.048},
   issn = {09214526},
   journal = {Physica B: Condensed Matter},
   month = {10},
   pages = {254-258},
   title = {Influence of annealing on spin pumping in sputtered deposited {Co/Pt} bilayer thin films},
   volume = {570},
   year = {2019}
}

@article{Kittel1948,
   author = {Charles Kittel},
   doi = {10.1103/PhysRev.73.155},
   issn = {0031-899X},
   issue = {2},
   journal = {Physical Review},
   month = {1},
   pages = {155-161},
   title = {On the Theory of Ferromagnetic Resonance Absorption},
   volume = {73},
   year = {1948}
}

@article{Arias1999,
  title = {Extrinsic contributions to the ferromagnetic resonance response of ultrathin films},
  author = {Arias, Rodrigo and Mills, D. L.},
  journal = {Phys. Rev. B},
  volume = {60},
  issue = {10},
  pages = {7395--7409},
  numpages = {0},
  year = {1999},
  month = {Sep},
  publisher = {American Physical Society},
  doi = {10.1103/PhysRevB.60.7395},
  url = {https://link.aps.org/doi/10.1103/PhysRevB.60.7395}
}

@article{Rossing1963,
   author = {Thomas D. Rossing},
   doi = {10.1063/1.1729582},
   issn = {0021-8979},
   issue = {4},
   journal = {Journal of Applied Physics},
   month = {4},
   pages = {995-995},
   title = {Resonance Linewidth and Anisotropy Variation in Thin Films},
   volume = {34},
   year = {1963}
}

@article{Heinrich1985,
   author = {B. Heinrich and J. F. Cochran and R. Hasegawa},
   doi = {10.1063/1.334991},
   issn = {0021-8979},
   issue = {8},
   journal = {Journal of Applied Physics},
   month = {4},
   pages = {3690-3692},
   title = {FMR linebroadening in metals due to two-magnon scattering},
   volume = {57},
   year = {1985}
}

@article{Yang2019,
   author = {Huaiwen Yang and Boyu Zhang and Xueying Zhang and Xi Yan and Wenlong Cai and Yinglin Zhao and Jirong Sun and Kang L. Wang and Dapeng Zhu and Weisheng Zhao},
   doi = {10.1103/PhysRevApplied.12.034004},
   issn = {2331-7019},
   issue = {3},
   journal = {Physical Review Applied},
   month = {9},
   pages = {034004},
   title = {Giant Charge-to-Spin Conversion Efficiency in {SrTiO$_3$} -Based Electron Gas Interface},
   volume = {12},
   year = {2019}
}

@article{Vaz2019,
   author = {Diogo C. Vaz and Paul Noël and Annika Johansson and Börge Göbel and Flavio Y. Bruno and Gyanendra Singh and Siobhan McKeown-Walker and Felix Trier and Luis M. Vicente-Arche and Anke Sander and Sergio Valencia and Pierre Bruneel and Manali Vivek and Marc Gabay and Nicolas Bergeal and Felix Baumberger and Hanako Okuno and Agnès Barthélémy and Albert Fert and Laurent Vila and Ingrid Mertig and Jean-Philippe Attané and Manuel Bibes},
   doi = {10.1038/s41563-019-0467-4},
   issn = {1476-1122},
   issue = {11},
   journal = {Nature Materials},
   month = {11},
   pages = {1187-1193},
   title = {Mapping spin–charge conversion to the band structure in a topological oxide two-dimensional electron gas},
   volume = {18},
   year = {2019}
}

@article{You2021,
   author = {Yunfeng You and Hiroto Sakimura and Takashi Harumoto and Yoshio Nakamura and Ji Shi and Cheng Song and Feng Pan and Kazuya Ando},
   doi = {10.1063/5.0035912},
   issn = {2158-3226},
   issue = {3},
   journal = {AIP Advances},
   month = {3},
   title = {Study of spin mixing conductance of single oriented {Pt} in {Pt}/{Ni$_8$$_1$Fe$_1$$_9$} heterostructure by spin pumping},
   volume = {11},
   year = {2021}
}

@article{Snchez2013,
   author = {J. C. Rojas Sánchez and L. Vila and G. Desfonds and S. Gambarelli and J. P. Attané and J. M. De Teresa and C. Magén and A. Fert},
   doi = {10.1038/ncomms3944},
   issn = {2041-1723},
   issue = {1},
   journal = {Nature Communications},
   month = {12},
   pages = {2944},
   title = {Spin-to-charge conversion using Rashba coupling at the interface between non-magnetic materials},
   volume = {4},
   year = {2013}
}

@article{Huihui2025,
author = {Ji, Huihui and Li, Minrui and Zhou, Guowei and Guo, Yiming and Gao, Xingguo and Zhou, Xuanchi and Liu, Liang and Xu, Xiaohong},
title = {Recent Progress On Low-Power Electrical Control of Magnetization in Transition Metal Oxide Heterostructures},
journal = {Advanced Functional Materials},
volume = {35},
number = {43},
pages = {2505227},
doi = {https://doi.org/10.1002/adfm.202505227},
url = {https://advanced.onlinelibrary.wiley.com/doi/abs/10.1002/adfm.202505227},
year = {2025}
}

@article{Trier2021,
   author = {Felix Trier and Paul Noël and Joo-Von Kim and Jean-Philippe Attané and Laurent Vila and Manuel Bibes},
   doi = {10.1038/s41578-021-00395-9},
   issn = {2058-8437},
   issue = {4},
   journal = {Nature Reviews Materials},
   month = {11},
   pages = {258-274},
   title = {Oxide spin-orbitronics: spin–charge interconversion and topological spin textures},
   volume = {7},
   year = {2021}
}

@article{Schoen2015,
  title = {Radiative damping in waveguide-based ferromagnetic resonance measured via analysis of perpendicular standing spin waves in sputtered permalloy films},
  author = {Schoen, Martin A. W. and Shaw, Justin M. and Nembach, Hans T. and Weiler, Mathias and Silva, T. J.},
  journal = {Phys. Rev. B},
  volume = {92},
  issue = {18},
  pages = {184417},
  numpages = {10},
  year = {2015},
  month = {Nov},
  publisher = {American Physical Society},
  doi = {10.1103/PhysRevB.92.184417},
  url = {https://link.aps.org/doi/10.1103/PhysRevB.92.184417}
}

@article{Pincus1960,
  title = {Excitation of Spin Waves in Ferromagnets: Eddy Current and Boundary Condition Effects},
  author = {Pincus, P.},
  journal = {Phys. Rev.},
  volume = {118},
  issue = {3},
  pages = {658--664},
  numpages = {0},
  year = {1960},
  month = {May},
  publisher = {American Physical Society},
  doi = {10.1103/PhysRev.118.658},
  url = {https://link.aps.org/doi/10.1103/PhysRev.118.658}
}
\clearpage \newpage

\beginsupplement
\title{\papertitle\linebreak Supplemental Material}
\maketitle
\begin{widetext}
    
\section{X-ray diffraction of KTO Substrate}
\begin{figure*}[htb]
    \centering
    \includegraphics[width=0.6\linewidth]{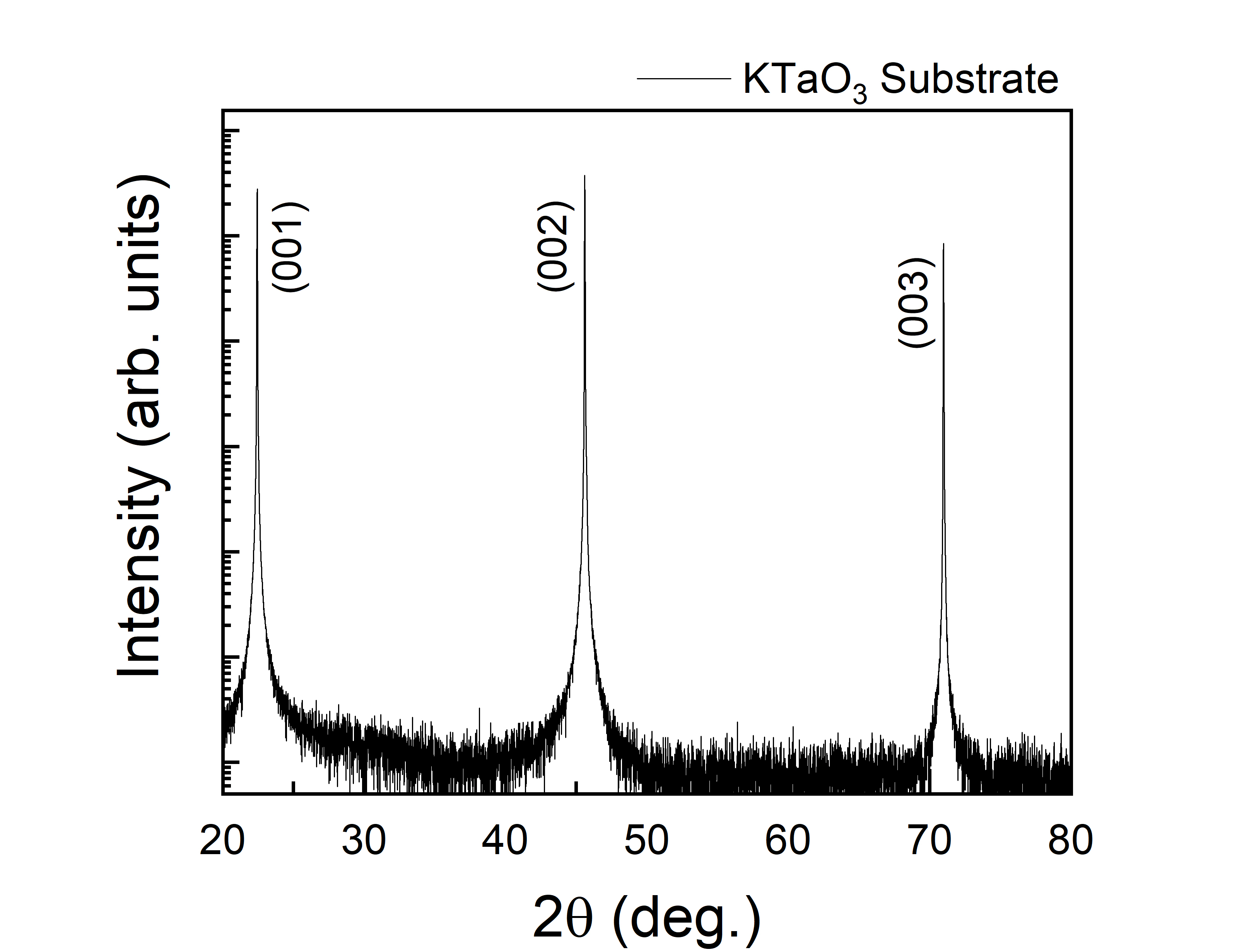}
    \caption{X-ray diffraction (XRD) of as-received single crystal KTaO$_3$ substrate}
\end{figure*}

Fig. S1 shows the x-ray diffraction results of the as-received single-crystal KTaO$_3$ substrate. The $\theta-2\theta$ XRD scan only shows (00$l$) peaks, confirming that the substrate is single-phase and (001) oriented with its c-axis perpendicular to the plane of the substrate.
\newpage
\section{The Gilbert damping coefficient estimation of all the samples}
Spin-pumping measurements were conducted on all the $\mathbf{\text{Ar}^{+}-\text{KTO}/\text{Py}}$ samples, each irradiated with $\text{Ar}^{+}$ ions for a different duration. Room-temperature Ferromagnetic Resonance (FMR) derivative absorption spectra ($\frac{dI(H)}{dH}$) were recorded over a microwave frequency range of $4$ to $10\ \text{GHz}$ (in $0.5\ \text{GHz}$ steps). The experimental data were analyzed using Lorentzian best fits (Equation 2), from which the resonance field ($H_r$) and linewidth ($\Delta H$) were precisely extracted. The full set of linewidth data ($\Delta H$ vs. frequency $f$) is summarized in Figure S2(b-f) for all samples. Finally, the effective damping constant ($\alpha_{\text{Ar}^{+}-\text{KTO}/\text{Py}}$) as shown in the inset was determined from the slope of the linear fit to the linewidth data using Equation 4, as detailed in the main text.
\begin{figure*}[htb]
    \centering
    \includegraphics[width=0.8\linewidth]{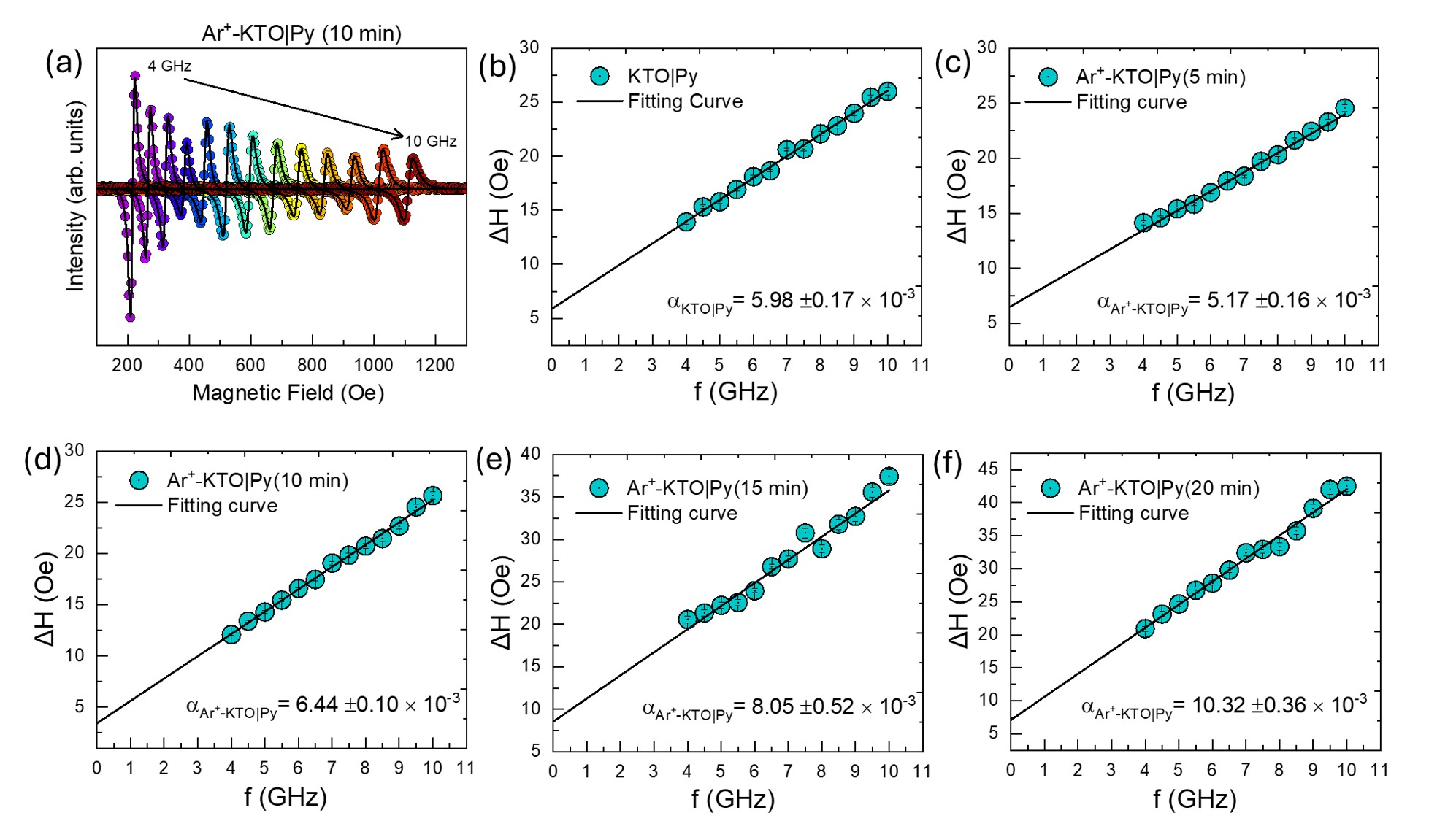}
    \caption{(a) Ferromagnetic Resonance (FMR) derivative absorption spectra ($\frac{dI(H)}{dH}$, where $I$ is the microwave absorption intensity) recorded at room temperature for microwave frequencies spanning from $4$ to $10\ \text{GHz}$ (in $0.5\ \text{GHz}$ increments). The experimental data is fitted with solid lines representing the Lorentzian best fits. (b–f) FMR linewidth ($\Delta H$) as a function of frequency for the non-irradiated KTO/Py reference bilayer and the $\text{Ar}^{+}$-irradiated $\text{Ar}^{+}-\text{KTO}/\text{Py}$ bilayers (for different irradiation durations). Solid lines indicate the theoretical fits using Equation 4 given in the main text.}
\end{figure*}
\newpage

\section{Microwave power dependence of the FMR spectra}
To ensure that the spin-pumping measurements were conducted within the linear excitation regime, we performed microwave power-dependent FMR measurements [Fig. S3(a)]. The FMR spectra were deconvoluted using the derivative Lorentzian line shape (Equation 2 in the main text), accounting for both symmetric and antisymmetric components. The symmetric component, which is primarily associated with the spin-pumping contribution, was extracted and analyzed across the investigated power range. As shown in Figure S3(b), the symmetric component amplitude ($k_s$) exhibits a clear linear dependence on the microwave power. This linear scaling confirms that the magnetization precession remains within the linear regime and that non-linear effects are negligible. These findings substantiate the reliability of our spin-pumping analysis and the accuracy of the extracted FMR parameters. 
\begin{figure*}[htb]
    \centering
    \includegraphics[width=1.0\linewidth]{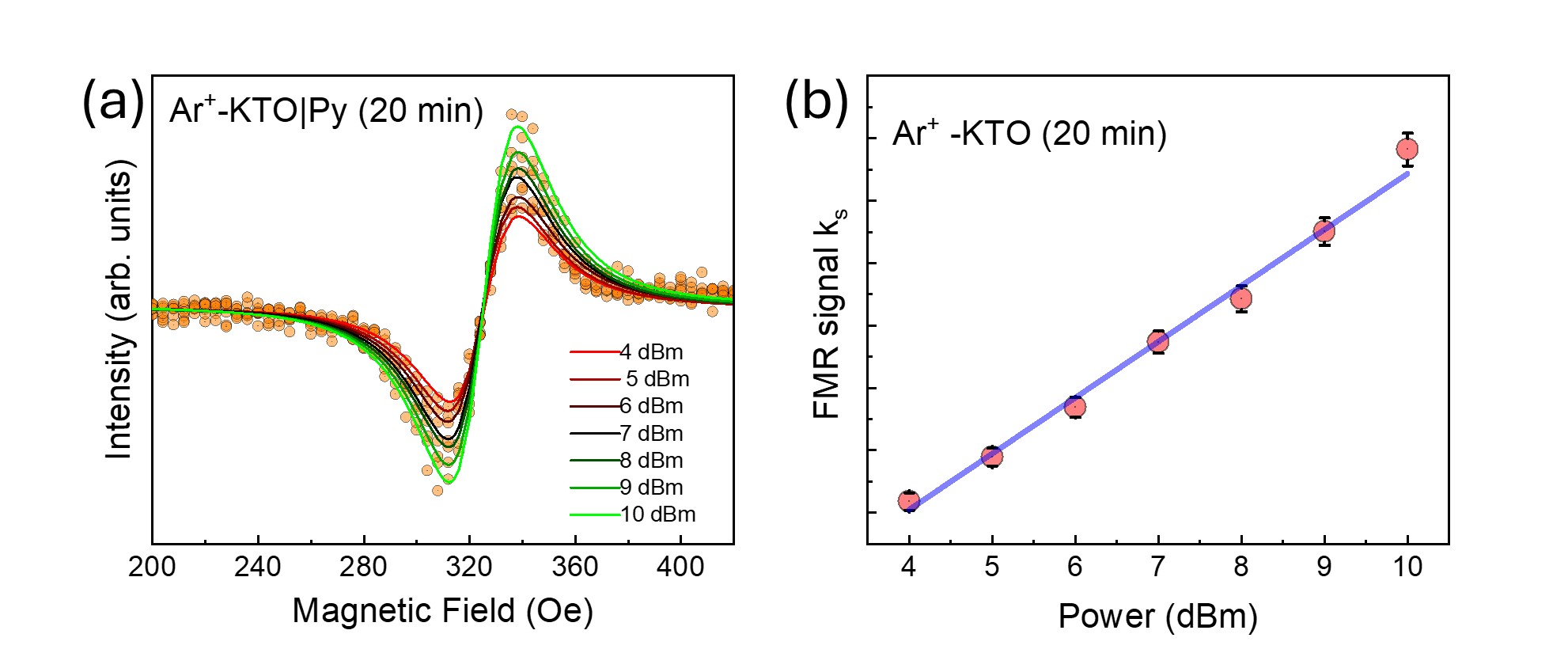}
    \caption{ (a) Microwave power dependence of the FMR spectra for a KTO sample irradiated with $Ar^{+}$ ions for 20 minutes. (b) Power dependence of the symmetric FMR signal amplitude ($k_{s}$), demonstrating a linear relationship.}
\end{figure*}

\newpage

\section{Extrinsic contributions to gilbert damping constant}
The magnetization dynamics in ferromagnet/heavy-metal (FM/HM) bilayers are governed by multiple damping channels. The total damping constant $\left(\alpha_{\text {tot }}\right)$ can be expressed as a superposition of the intrinsic Gilbert damping and various extrinsic contributions, as follows
$$
\alpha_{t o t}=\alpha_{i n t}+\alpha_{r a d}+\alpha_{e d d y}+\alpha_{s p}
$$

Here, $\alpha_{\text {int }}$ represents the intrinsic Gilbert damping constant, while $\alpha_{s p}$ denotes the additional damping contribution arising from spin pumping into the adjacent layer. The terms $\alpha_{\text {rad }}$ and $\alpha_{\text {eddy }}$ account for extrinsic contributions originating from radiative and eddy-current damping, respectively.

In coplanar waveguide (CPW)-based FMR, radiative damping originates from energy dissipation via inductive coupling between the precessing magnetization and the waveguide transmission line. Additionally, in metallic ferromagnets, time-varying magnetic flux induces eddy currents that exert a restorative torque on the magnetization dynamics, providing an extra damping channel. We have analytically calculated these extrinsic contributions specifically for our experimental geometry and sample dimensions.

For uniform magnetization precession, the radiative damping contribution is given by the expression\cite{Schoen2015}
$$
\alpha_{r a d}=\frac{\eta \gamma \mu_{o}^{2} M_{s} \delta l}{2 Z_{o} w}
$$
where $\gamma$ is the gyromagnetic ratio, $Z_{o}$ is the waveguide impedance, $w(=250~\mu \mathrm{m})$ denotes the width of the CPW center conductor, $\delta$ represents the thickness, and $l$ is the length of the sample. The dimensionless parameter $\eta$ accounts for the FMR mode profile, which is taken as 0.25 for the uniform mode. Based on our specific experimental geometry, the calculated value for radiative damping is $\alpha_{\text {rad }}=3.94 \times 10^{-5}$.

Similarly, the eddy-current damping coefficient, $\alpha_{\text {eddy }}$, was evaluated using the following expression\cite{Pincus1960,Schoen2015}
$$
\alpha_{e d d y}=\frac{C \gamma \mu_{o}^{2} M_{s} \delta^{2}}{16 \rho_{x x}}
$$
where $\mu_{0}$ is the vacuum permeability, $\gamma$ is the gyromagnetic ratio, $M_{s}$ is the saturation magnetization, and $\delta$ denotes the film thickness. The parameter $\rho_{x x}$ represents the longitudinal resistivity of the sample, while $C$ is a correction factor describing the eddy-current distribution, taken as 0.5 for the uniform FMR mode. For our specific geometry and sample parameters, the calculated value of $\alpha_{\text {eddy }}=3.48 \times 10^{-6}$.

The calculated values for both radiative and eddy-current damping are several orders of magnitude smaller than the intrinsic and spin-pumping contributions ( $\alpha_{A r^{+}-K T O / P y} \approx 1.03 \times 10^{-2}$ ). Consequently, these extrinsic factors are negligible and do not significantly influence the total damping analysis presented in the main manuscript.

\end{widetext}

\end{document}